\documentclass[a4paper]{amsart}

\usepackage{graphicx}

\theoremstyle{plain} 
\newtheorem{thm}{Theorem}

\newtheorem{rem}[thm]{Remark}

\newcommand{\gr}[1]{\boldsymbol{#1}}
\newcommand{\m}{\gr{m}}
\newcommand{\n}{\gr{n}}
\newcommand{\s}{\gr{s}}

\newcommand{\F}{\mathsf{F}}
\newcommand{\C}{\mathsf{C}}
\newcommand{\I}{\mathsf{I}}
\newcommand{\T}{\mathsf{T}}

\renewcommand{\P}{\mathsf{P}}
\newcommand{\tr}{\operatorname{tr}}

\newcommand{\de}[2]{\frac{\partial #1}{\partial #2}}
\newcommand{\qRq}{\quad\Rightarrow\quad}
\newcommand{\tens}{\mathsf}
\newcommand{\dde}[2]{\frac{\partial^2 #1}{\partial#2}}

\begin{document}

\title[Comparison between active strain and active stress]{A
  comparison between active strain and active stress in transversely isotropic
  hyperelastic materials}

\author{Giulia Giantesio, Alessandro Musesti}
\address[Giulia Giantesio, Alessandro Musesti]{Dipartimento di Matematica e Fisica ``N. Tartaglia'', Universit\`a
  Cattolica del Sacro Cuore, via dei Musei 41, 25121 Brescia, Italy}
\email{giulia.giantesio@unicatt.it,alessandro.musesti@unicatt.it}

\author{Davide Riccobelli}
\address[Davide Riccobelli]{MOX, Dipartimento di Matematica,
Politecnico di Milano, Via Bonardi 9, 20133 Milano, Italy}
\email{davide.riccobelli@polimi.it}

\date{\today}



\begin{abstract}
 Active materials are media for which deformations can
  occur in absence of loads, given an external stimulus. Two
  approaches to the modeling of such materials are mainly used in
  literature, both based on the introduction of a new tensor: an
  additive stress $\P_\text{act}$ in the \emph{active stress} case and
  a multiplicative strain $\F_a$ in the \emph{active strain} one.
  Aim of this paper is the comparison between the two approaches on
  simple shears.
  
  Considering an incompressible and transversely isotropic material,
  we design constitutive relations for $\P_\text{act}$ and $\F_a$ so
  that they produce the same results for a uniaxial deformation along
  the symmetry axis. We then study the two approaches in the case of a
  simple shear deformation. In a hyperelastic setting, we show that
  the two approaches produce different stress components along a
  simple shear, unless some necessary conditions on the strain energy
  density are fulfilled. However, such conditions are very restrictive
  and rule out the usual elastic strain energy functionals. Active stress
  and active strain therefore produce different results in shear, even
  if they both fit uniaxial data.

  Our results show that experimental data on the stress-stretch
  response on uniaxial deformations are not enough to establish which
  activation approach can capture better the mechanics of active
  materials. We conclude that other types of deformations, beyond the
  uniaxial one, should be taken into consideration in the modeling of
  such materials.
\end{abstract}

\maketitle

\section{Introduction}
The main feature of a body made of an active material is the ability
of changing its mechanical properties by an external stimulus (for
example an electrical signal in muscles). During the last decades,
many efforts have been made in order to study the properties of active
materials, from smart materials, such as dielectric elastomers to
biological ones, such as muscles and cardiac tissue. Needless to say,
the technological applications of such materials are copious and a
good modeling of biological active tissues can be very helpful to
biomedical sciences.

Two different mathematical approaches are largely used in the
literature for modeling \emph{activation} \cite{AmbPez}: in the most
popular one, named \emph{active stress}, an extra term $\P_\text{act}$
is added to the stress accounting for the contribution given by the
activation (see for example \cite{martins,blemker,thomas}). On the
contrary, the \emph{active strain} approach, firstly proposed by
Kondaurov and Nikitin~\cite{kond87} and then developed by Taber and
Perucchio \cite{taber} in the modeling of cardiac tissue, was inspired
by classical ideas in plasticity and previous theories of growth and
morphogenesis; the key ingredient is a multiplicative decomposition
$\F=\F_e\F_a$ of the deformation gradient, where $\F_a$ is the
activation distortion and $\F_e$ accounts for the storage of elastic
energy \cite{NarTer}. Both approaches have strong motivations: for
instance, in the case of muscle tissue the active stress approach can
easily fit to experiments, while the active strain approach is much
more inherent to the mechanism of contraction of sarcomeres, the
so-called sliding filament theory.

Aim of the present paper is to show that active stress and active
strain give different stress components on a simple shear
deformation, even if they make the same predictions on uniaxial
elongations. Our results show that experimental data on the
stress-stretch response on uniaxial deformations are not enough to
establish which activation approach can capture better the activation
mechanics. A further study of other deformations, such as simple
shears, would be important in order to develop a realistic model of an
active material.

A comparison between the two approaches has been previously addressed
from other points of view
\cite{AmbPez,Rossi2012,heidlauf2013treatment}; here we present a
broader study in the case of a general hyperelastic material
(Sect.~\ref{sec:stressfromstrain}) and perform a quantitative analysis
in the case of a fiber-reinforced Mooney-Rivlin material
(Sect.~\ref{sec:Fainco}) and of a material with an exponential energy
typically used in the modeling of skeletal muscle tissue
(Sect.~\ref{sec:energyebi}). Such a comparison can be very important
in the choice of which approach one should use in the modeling of
activation, especially when shear deformations are involved.  In
Sect.~\ref{sec:fr} we analyze the special case of fiber-reinforced
materials.

Considering a passive material which is hyperelastic, transversely
isotropic and incompressible, we proceed in this way: given a strain
energy functional we consider a constant active strain. We design the
active stress so that the predictions coincide for uniaxial
deformations along the material symmetry axis. Then, we compare the
two activation models on a simple shear deformation. It turns out that
the stresses corresponding to the two activation approaches are considerably
different, unless the energy satisfies a very restrictive condition.

In the choice of the form of the active terms we follow the common
assumptions used in the literature about active transversely isotropic
materials. Namely, in the active stress approach we assume
$\P_\text{act}$ to depend only on the stretch in the direction of
anisotropy, whereas in the active strain approach we consider $\F_a$ as an
incompressible contraction along that direction.

In Sect.~\ref{sec:energyebi} we analyze a more complex energy related
to skeletal muscle tissue. Here the active stress is computed from
experimental data along a uniaxial deformation and the active strain
depends on the stretch along the muscle fibers.  Again, the two
approaches give very different stress components on simple shears. This has important consequences for the modelling of muscles when deformations other than the uniaxial extension are involved. For instance, the deformation of a pennate muscle, where the
muscle fibers are attached obliquely to the tendon, is definitely not
a uniaxial deformation along the fibers, and also the cross-fiber
simple shear plays an important role.

Finally, we note that a few other activation approaches are proposed
in the literature, see for instance~\cite{ebi,Hernandez13,Pae15}. In
Sect.~\ref{sec:decoupled} we discuss one of them which is typically used for
fiber-reinforced materials, where the active strain decomposition is
applied only to the anisotropic part of the elastic energy. We call
such an approach \emph{decoupled active strain}. Under some mild
assumptions, we show that decoupled active strain is
completely equivalent to the addition of an active stress.

\section{Hyperelastic activation}\label{sec:pre}

The goal of this section is to introduce the \emph{active
  strain} and \emph{active stress} methods used to model
the activation of a material. Before illustrating these approaches, we
stipulate the following general assumptions.

We consider a passive material which is hyperelastic and
incompressible with a strain energy density $W_\text{pas}(\F)$, where $\F$ is the deformation gradient. The
first Piola-Kirchhoff stress tensor writes
\begin{equation}\label{eq:Pp}
\P_\text{pas}(\F)=\de{W_\text{pas}}{\F}(\F)-p\F^{-\T}, 
\end{equation}
where $p$ is a Lagrange multiplier enforcing the incompressibility
constraint 
\[
J:=\det\F=1.
\]
Moreover we assume that the material is frame-indifferent and
transversely isotropic with structural tensor $\m\otimes \m$, where
$\m$ is the direction of anisotropy in the reference configuration (for
instance, the direction of the fibers in the case of skeletal
muscles). Then, it is well-known that the elastic energy may be formulated as a function
of the following five invariants of the right Cauchy-Green deformation
tensor $\C=\F^\T\F$:
\[
I_1=\tr\C,\quad I_2=\frac 1 2 \left(I_1^2-\tr \C^2\right),\quad
I_3=\det\C,\quad I_4=\m\cdot\C\m,\quad I_5=\m\cdot\C^2\m.
\]

Concerning activation, we briefly recall the two main approaches.

\null

\begin{figure}
\begin{center}
\includegraphics[scale=.7]{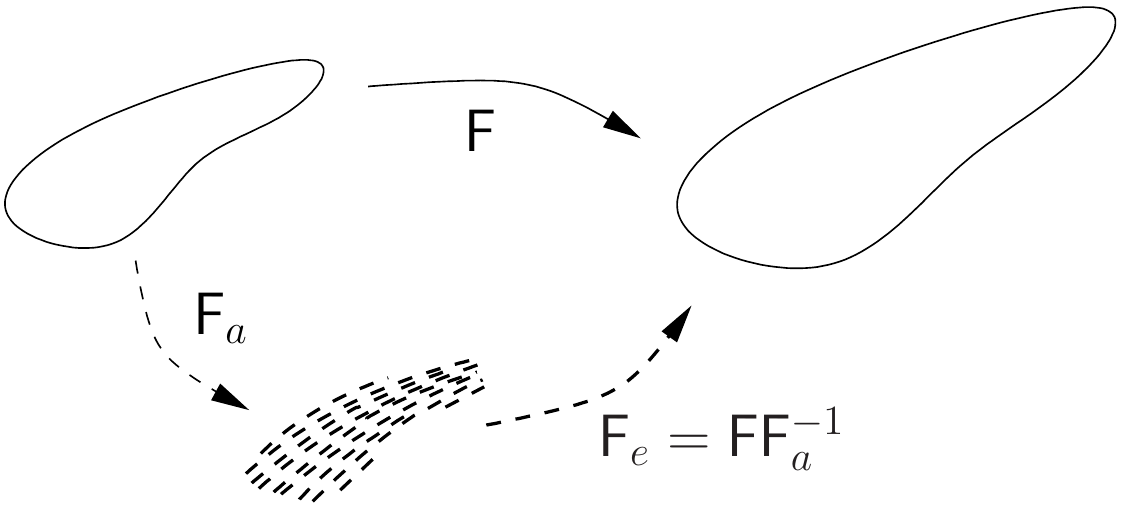}
\end{center}
\caption{A pictorial representation of the Kr\"oner-Lee decomposition.} 
\label{fig:active}
\end{figure}

\noindent\textbf{Active strain:} using the so called Kr\"oner-Lee
decomposition in the theory of elastoplasticity (see
Fig.~\ref{fig:active}), we factorize the deformation gradient as
\[
\F=\F_e\F_a,
\]
where $\F_a$ has to be constitutively provided. The tensors $\F_a$ and
$\F_e=\F\F_a^{-1}$ are named \emph{active strain} and \emph{elastic
  strain}, respectively. Notice that $\F_a$ and $\F_e$ may not be the
gradients of some deformation.
The active strain $\F_a$ represents a virtual distortion of the
relaxed configuration due to activation and only the tensor $\F_e$ is
responsible for the storing of elastic energy. Then the strain energy
density of the active material is given by
\begin{equation}
\label{eq:Wstrain}
W_\text{strain}(\F;\F_a)=(\det\F_a)W_\text{pas}(\F\F_a^{-1}),
\end{equation}
see for instance~\cite{kond87,taber,NarTer}, and the stress tensor writes
\begin{equation}
  \label{eq:astrain}
\P_\text{strain}(\F;\F_a)=(\det\F_a) \P_\text{pas}(\F\F_a^{-1})\F_a^{-\T},
\end{equation}
where $\P_\text{pas}$ is given by \eqref{eq:Pp}. 

In the active strain approach the energy density $W_\text{strain}$ inherits
the same mathematical properties of $W_\text{pas}$; for instance, the
polyconvexity of the latter ensures the same regularity of the
former \cite{neff2003some}.

In the following sections, we consider an isochoric active strain tensor of the form
\begin{equation}
\label{eq:Fa}
\F_a=
(1-a) \m\otimes\m+\frac{1}{\sqrt{1-a}}(\I-\m\otimes\m),
\qquad 0\leq a < 1.
\end{equation}
Such a choice, which is customary in the literature (see for instance
\cite{taber,pezzuto2014,gammaderiv}), allows us to obtain the whole
tensor $\F_a$ by means of a single scalar parameter $a$, accounting
for the contraction of the material along the symmetry direction
$\m$. In the literature, also the case of a non-isochoric
$\F_a$ has been considered
\cite{ambrosi2011electromechanical,NarTer,noi}, however such a
constitutive choice is less popular and will not be taken into
consideration in this paper.

\null
 
\noindent\textbf{Active stress:} we additively decompose the total stress as
\begin{equation}
\label{eq:activestressgenerale}
\P_\text{stress}(\F)=\P_\text{pas}(\F) + \P_\text{act}(\F)
\end{equation}
where $\P_\text{act}$, to be constitutively provided, is the
stress due to the activation (see for
instance~\cite{Odegard2008,Path2010,heid2013}).

The formulation given in \eqref{eq:activestressgenerale} is quite general. In principle, if we set
\[
\P_\text{act}(\F) = (\det\F_a) \P_\text{pas}(\F\F_a^{-1})\F_a^{-\T} - \P_\text{pas}(\F),
\]
then by a suitable choice of the active stress one can recover the active strain approach.
However, in the literature the active stress $\P_\text{act}$ has to
fulfill some modeling prescriptions, which are often incompatible with
such a choice.

In the case of transversely isotropic
materials with direction of anisotropy $\m$, it is
usually assumed that $\P_\text{act}$ depends on $\F$ only through the
pseudo-invariant
\[
I_4=\F\m\cdot\F\m
\]
in the following way:
\begin{equation}
  \label{eq:Pact}
\P_\text{act}(\F)=2S(\sqrt{I_4})\F\m\otimes\m,
\end{equation}
where $S$ is a scalar function. 
One may notice that, according to \eqref{eq:Pact}, the
non-null components of $\P_\text{act}$ are all along
$\F\m\otimes\m$ while in the active strain approach all the components are involved.

Denoting by $W_\text{act}$ a
primitive function of $S$, one has that
\begin{equation}
  \label{eq:astress}
\P_\text{stress}=\de{W_\text{pas}}{\F}(\F)+\frac{W_\text{act}'(\sqrt{I_4})}{\sqrt{I_4}}\F\m\otimes\m
-p_\text{stress}\F^{-\T}.
\end{equation}

\begin{rem}
It is important to note that
$W_\text{act}$ should not be
physically interpreted as a strain energy density, but only as a primitive
function of the active stress. In any case, from a mathematical viewpoint one can define
\begin{equation}
\label{eq:stresscons}
W_\text{stress}(\F)=W_\text{pas}(\F)+W_\text{act}(\sqrt{I_4}).
\end{equation}

The function $W_\text{act}$ can affect the mathematical properties of
the total energy, as discussed for instance in~\cite{Path2010}. The
polyconvexity or the rank-one convexity of the total energy are no
more ensured, even if $W_\text{pas}$ is convex.
\end{rem}

In the next sections we compare the two activation approaches,
namely active strain and active stress. We focus on two families
of homogeneous deformations: the uniaxial deformation along the
direction of anisotropy $\m$ and the simple shear orthogonal to $\m$.
Such a shear modifies the elongation of the body in the direction of
anisotropy, so that it allows us to point at differences between the
two approaches.

\begin{figure}
\begin{center}
\includegraphics[scale=.7]{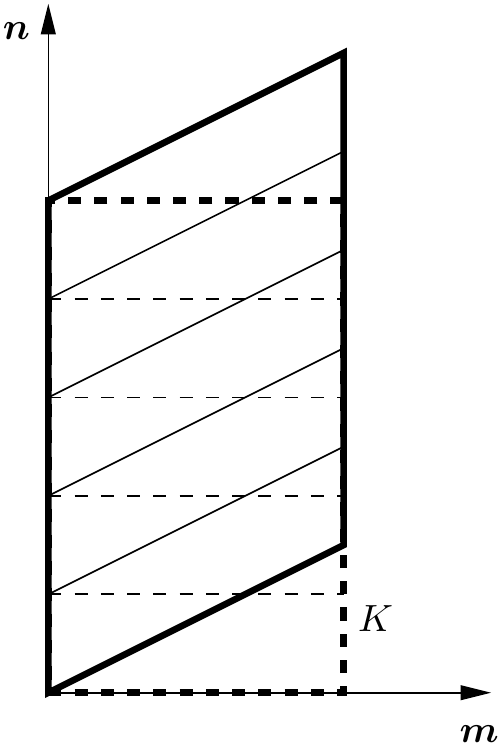}
\end{center}
\caption{Pictorial representation in the plane of $\m,\n$ of the simple
  shear~\eqref{eq:shear}.} 
\label{fig:shear}
\end{figure}

Specifically, we consider the uniaxial incompressible deformation gradient
\begin{equation}
\label{eq:uniax}
\F_\lambda=
\lambda \m\otimes\m+\frac{1}{\sqrt{\lambda}}(\I-\m\otimes\m),
\end{equation}
and, given a direction $\n$
orthogonal to $\m$, the simple shear deformation whose gradient is given by
\begin{equation}
\label{eq:shear}
\F_K=\I+K\n\otimes\m,
\end{equation}
where
$K$ is the amount of shear
(Fig.~\ref{fig:shear}). Notice that in the first case
the stretch along the preferred direction $\m$ is given by $\lambda$,
while in the simple shear it is given by $\sqrt{1+K^2}$.

\section{Comparing active stress and active strain on a simple shear}\label{sec:stressfromstrain} 

In this section we consider an active strain $\F_a$ of the
form~\eqref{eq:Fa} and an active stress $\P_a$ of the
form~\eqref{eq:Pact}. Starting from the same passive elastic energy
density and imposing that the two activation approaches coincide on
uniaxial deformations along the direction of anisotropy, we
will compare them on a simple shear.

We consider a homogeneous elastic strain energy density for a
transversely isotro\-pic incompressible material of the form
\[
W(I_1,I_2,I_4,I_5).
\]
Then we study the response of the two activation approaches on the uniaxial deformation $\F_\lambda$ \eqref{eq:uniax} 
and on the simple shear $\F_K$ \eqref{eq:shear}, 
while we denote:
\begin{align*}
&\F_{\lambda   e}=\F_\lambda\F_a^{-1},\qquad \F_{K e}=\F_K\F_a^{-1}.
\end{align*}

Moreover, let us introduce the notation
\begin{align*}
&\C_e=\F^{\tens T}_e\F_e,\quad\C_\lambda=\F_\lambda^{\tens T}\F_\lambda,\quad
\C_{\lambda e}=\F_{\lambda e}^{\tens T}\F_{\lambda   e},\\
&\C_K=\F^{\tens T}_K\F_K,\quad
\C_{Ke}=\F_{K e}^{\tens T}\F_{K e}.
\end{align*}
Then, by~\eqref{eq:Wstrain} and recalling that $\det\F_a=1$, the energy density of the material activated with the active strain
approach is given by
\begin{equation}
\label{eq:estrainl}
W_\text{strain}=W\Big(I_1(\C_{\lambda e}),I_2(\C_{\lambda e}),I_4(\C_{\lambda e}),I_5(\C_{\lambda e})\Big).
\end{equation}

On the other hand, by~\eqref{eq:stresscons} the energy density of the
material activated with the active stress approach has the form
\begin{equation}
\label{eq:estress}
W_\text{stress}=W\Big(I_1(\C_{\lambda}),I_2(\C_{\lambda}),I_4(\C_{\lambda}),I_5(\C_{\lambda})\Big)+W_\text{act}\left(\sqrt{I_4(\C_{\lambda})}\right).
\end{equation}

Now we want to find $W_\text{act}$ such that the two energy
densities~\eqref{eq:estrainl} and~\eqref{eq:estress} coincide on the
deformation $\F_\lambda$ for any $\lambda$ and any given value
of the activation parameter $a$. Hence we have to choose
\begin{multline}
\label{eq:Wact}
W_\text{act}(\lambda;a)=W\Big(I_1(\C_{\lambda e}),I_2(\C_{\lambda e}),I_4(\C_{\lambda e}),I_5(\C_{\lambda e})\Big)
\\
-W\Big(I_1(\C_{\lambda}),I_2(\C_{\lambda}),I_4(\C_{\lambda}),I_5(\C_{\lambda})\Big),
\end{multline}
where we pointed out the dependence of
$W_\text{act}$ on the amount of stretch and on the activation
parameter.

For a general deformation $\F$, the elastic energy density
corresponding to the active strain model will be directly computed
using $W(I_1(\C_e),I_2(\C_e),I_4(\C_e),I_5(\C_e))$.  On the contrary, the elastic energy
density corresponding to the active stress model will be given by
\[
W(I_1,I_2,I_4,I_5)+W_\text{act}(\sqrt{I_4};a),
\] 
where the function $W_\text{act}$ is given in~\eqref{eq:Wact}.

\null

We now consider the simple shear deformation $\tens{F}_K$ given
by~\eqref{eq:shear}. In this case we have $\sqrt{I_{4}(\C_K)}=\sqrt{1+K^2}$ and
{\footnotesize
\begin{align*}
&I_1(\C_{Ke})=\frac{1+K^2}{(1-a)^2}+2(1-a),\quad
&&I_2(\C_{Ke})=\frac{2+K^2}{1-a}+(1-a)^2,\\
&I_4(\C_{Ke})=\frac{1+K^2}{(1-a)^2},\quad
&&I_5(\C_{Ke})=\frac{(1+K^2)^2}{(1-a)^4}+\frac{K^2}{1-a},\\
&I_1(\C_K)=3+K^2, \quad
&&I_2(\C_K)=3+K^2,\\
&I_4(\C_K)=1+K^2,\quad
&&I_5(\C_K)=K^2+(1+K^2)^2,\\
&I_1(\C_{\lambda e})=\frac{1+K^2}{(1-a)^2}+2\frac{1-a}{\sqrt{1+K^2}},\quad
&&I_2(\C_{\lambda e})=\frac{(1-a)^2}{1+K^2}+2\frac{\sqrt{1+K^2}}{1-a},\\
&I_4(\C_{\lambda e})=\frac{1+K^2}{(1-a)^2},\quad
&&I_5(\C_{\lambda e})=\frac{(1+K^2)^2}{(1-a)^4},\\
&I_1(\C_\lambda)=1+K^2+\frac{2}{\sqrt{1+K^2}},\quad
&&I_2(\C_\lambda)=\frac{1}{1+K^2}+2\sqrt{1+K^2},\\
&I_4(\C_\lambda)=1+K^2,\quad
&&I_5(\C_\lambda)=(1+K^2)^2.
\end{align*}
}

Imposing that active strain and active stress have the same energy density (and
hence the same stress tensor field) both on every uniaxial deformation
$\tens{F}_\lambda$ and every simple shear $\tens{F}_K$, then
\begin{multline*}
W\Big(I_1(\C_{Ke}),I_2(\C_{Ke}),I_4(\C_{Ke}),I_5(\C_{Ke})\Big)\\
=W(I_1(\C_K),I_2(\C_K),I_4(\C_K),I_5(\C_{K}))+W_\text{act}(\sqrt{1+K^2};a),
\end{multline*}
where the function $W_\text{act}$ is given in~\eqref{eq:Wact}.

Hence for every $0\leq a<1$ and $K\geq 0$ one has
\begin{align*}
&W\left(\frac{1+K^2}{(1-a)^2}+2(1-a),\frac{2+K^2}{1-a}+(1-a)^2,\frac{1+K^2}{(1-a)^2},
\frac{(1+K^2)^2}{(1-a)^4}+\frac{K^2}{1-a}\right)\\
-&W\left(\frac{1+K^2}{(1-a)^2}+2\frac{1-a}{\sqrt{1+K^2}},\frac{(1-a)^2}{1+K^2}+2\frac{\sqrt{1+K^2}}{1-a},\frac{1+K^2}{(1-a)^2},\frac{(1+K^2)^2}{(1-a)^4}\right)\\
=&W\left(3+K^2,3+K^2,1+K^2,K^2+(1+K^2)^2\right)\\
-&W\left(1+K^2+\frac{2}{\sqrt{1+K^2}},\frac{1}{1+K^2}+2\sqrt{1+K^2},1+K^2,
(1+K^2)^2\right).
\end{align*}
Setting for convenience $1+K^2=\ell^2$ and $\frac{1}{1-a}=x$, the equation becomes
\begin{equation}
  \label{eq:I5}
  \begin{aligned}
&W\left({\ell^2}{x^2}+\frac2x,(1+\ell^2)x+\frac{1}{x^2},{\ell^2}{x^2},
{\ell^4}{x^4}+(\ell^2-1){x}\right)\\
-&W\left({\ell^2}{x^2}+\frac{2}{\ell x},\frac{1}{\ell^2 x^2}+{2\ell}{x},{\ell^2}{x^2},{\ell^4}{x^4}\right)\\
=&W\left(2+\ell^2,2+\ell^2,\ell^2,\ell^4+\ell^2-1\right)
-W\left(\ell^2+\frac{2}{\ell},\frac{1}{\ell^2}+2\ell,\ell^2,
\ell^4\right).
\end{aligned}
\end{equation}
Differentiating w.r.t. $\ell$ and letting $\ell\to 1$ one gets
\begin{align*}
&\de{W}{I_1}\left(x^2+\frac 2 x,\frac{1}{x^2}+2x,
x^2,x^4\right)\\
+\frac 1 x &\de{W}{I_2}\left(x^2+\frac 2 x,\frac{1}{x^2}+2x,
x^2,x^4\right)\\
+x^2&\de{W}{I_5}\left(x^2+\frac 2 x,\frac{1}{x^2}+2x,
x^2,x^4\right)\\
=x&\left(\de{W}{I_1}(3,3,1,1)+\de{W}{I_2}(3,3,1,1)+
\de{W}{I_5}(3,3,1,1)\right).
\end{align*}
On the other hand, differentiating w.r.t. $x$ and letting $x\to 1$ one gets
\begin{align*}
2(\ell^2-1)\ell^2&\de{W}{I_1}\left(2+\ell^2,2+\ell^2,\ell^2,\ell^4+\ell^2-1\right)\\
+(\ell^2-1)\ell^2&\de{W}{I_2}\left(2+\ell^2,2+\ell^2,\ell^2,\ell^4+\ell^2-1\right)\\
+2\ell^4&\de{W}{I_4}\left(2+\ell^2,2+\ell^2,\ell^2,\ell^4+\ell^2-1\right)\\
+\ell^2(4\ell^4+\ell^2-1)&\de{W}{I_5}\left(2+\ell^2,2+\ell^2,\ell^2,\ell^4+\ell^2-1\right)\\
=2\ell(\ell^3-1)&\de{W}{I_1}\left(\ell^2+\frac{2}{{\ell}},\frac{1}{\ell^2}+2{\ell},\ell^2,
\ell^4\right)
\\
+2(\ell^3-1)&\de{W}{I_2}\left(\ell^2+\frac{2}{{\ell}},\frac{1}{\ell^2}+2{\ell},\ell^2,
\ell^4\right)
\\
+2\ell^4&\de{W}{I_4}\left(\ell^2+\frac{2}{{\ell}},\frac{1}{\ell^2}+2{\ell},\ell^2,
\ell^4\right)
\\
+4\ell^6&\de{W}{I_5}\left(\ell^2+\frac{2}{{\ell}},\frac{1}{\ell^2}+2{\ell},\ell^2,
\ell^4\right).
\end{align*}

By taking the mixed second derivative of~\eqref{eq:I5} and letting
both $x\to 1$ and $\ell\to 1$, one gets
\begin{multline*}
\de{W}{I_1}\left(3,3,1,1\right)+2\de{W}{I_2}\left(3,3,1,1\right)-\de{W}{I_5}\left(3,3,1,1\right)\\
=2\dde{W}{I_1 I_4}\left(3,3,1,1\right)+4\dde{W}{I_1 I_5}\left(3,3,1,1\right)+2\dde{W}{I_2 I_4}\left(3,3,1,1\right)\\
+4\dde{W}{I_2 I_5}\left(3,3,1,1\right)+2\dde{W}{I_4 I_5}\left(3,3,1,1\right)+4\dde{W}{I_5^2}\left(3,3,1,1\right)
\end{multline*}
which gives a very particular relation for the elastic moduli in the
identity. For instance, imposing the previous condition on the energy
density~\eqref{eq:Webi} considered in Sect.~\ref{sec:energyebi},
which depends also on $I_5$, we get the necessary condition
\[
2(\alpha+2\beta)w_0^2-(2\alpha+10\beta+3)w_0+6(\beta+1)
=0
\]
which holds only for very special values of the constitutive parameters.

Even if we drop out the dependence of the energy on $I_5$, as we will do in the sequel, from \eqref{eq:I5} we have 
\begin{multline}
\label{eq:Wlx}
W\left(\ell^2x^2+\frac 2 x,(1+\ell^2)x+\frac{1}{x^2},\ell^2x^2\right)
-W\left(\ell^2x^2+\frac{2}{\ell x},\frac{1}{\ell^2x^2}+2\ell x,\ell^2x^2\right)\\
=W\left(2+\ell^2,2+\ell^2,\ell^2\right)
-W\left(\ell^2+\frac{2}{\ell},\frac{1}{\ell^2}+2\ell,\ell^2\right),
\end{multline}
for every $x\geq 1$ and $\ell\geq 1$.

Condition~\eqref{eq:Wlx} results to be very restrictive and rules out
any typical energy density used for elastic materials. Indeed, the
only elastic energy density that we have found to satisfy the
equivalence between active stress and active strain both on
$\F_\lambda$ and on $\F_K$ is
\[
  W(I_1,I_4)=cI_1\sqrt{I_4}+f(I_4).
\]
In such a case, eq.~\eqref{eq:Wlx} is satisfied for any $x,\ell\geq
1$. However, that energy has a very particular form and we are not
aware of any model of nonlinear elasticity where it is used.

\subsection{The fiber-reinforced case}
\label{sec:fr}
Among the transversely isotropic media, an important role is played by
the so-called \emph{fiber-reinforced materials}, for which the strain
energy density splits as a sum of isotropic and anisotropic
contributions. For the sake of simplicity, we will assume that the
anisotropic term does not depend on $I_5$, so that
\[
W=W_\text{iso}(I_1,I_2)+W_\text{aniso}(I_4).
\]
Then~\eqref{eq:Wlx} becomes
\begin{multline}
\label{eq:Wlxreinforced}
W_\text{iso}\left(\ell^2x^2+\frac 2 x,(1+\ell^2)x+\frac{1}{x^2}\right)
-W_\text{iso}\left(\ell^2x^2+\frac{2}{\ell x},\frac{1}{\ell^2x^2}+2\ell x\right)\\
=W_\text{iso}\left(2+\ell^2,2+\ell^2\right)
-W_\text{iso}\left(\ell^2+\frac{2}{\ell},\frac{1}{\ell^2}+2\ell\right),
\end{multline}
for every $x\geq 1$ and $\ell\geq 1$.

Differentiating w.r.t. $\ell$ and letting $\ell=1$ one gets
\begin{multline}
\label{eq:difflreinforced}
\de{W_\text{iso}}{I_1}\left(x^2+\frac 2 x,\frac{1}{x^2}+2x\right)
+\frac 1 x \de{W_\text{iso}}{I_2}\left(x^2+\frac 2 x,\frac{1}{x^2}+2x\right)\\
=x\left(\de{W_\text{iso}}{I_1}(3,3)+\de{W_\text{iso}}{I_2}(3,3)\right).
\end{multline}
On the other hand, differentiating w.r.t. $x$ and letting $x=1$ one
gets
\begin{multline}
\label{eq:diffxreinforced}
2(\ell^2-1)\ell^2\de{W_\text{iso}}{I_1}\left(2+\ell^2,2+\ell^2\right)
+(\ell^2-1)\ell^2\de{W_\text{iso}}{I_2}\left(2+\ell^2,2+\ell^2\right)\\
=2\ell(\ell^3-1)\de{W_\text{iso}}{I_1}\left(\ell^2+\frac{2}{{\ell}},\frac{1}{\ell^2}+2{\ell}\right)
+2(\ell^3-1)\de{W_\text{iso}}{I_2}\left(\ell^2+\frac{2}{{\ell}},\frac{1}{\ell^2}+2{\ell}\right).
\end{multline}
By taking the mixed (second) derivative of~\eqref{eq:Wlx} and letting both $x=1$ and
$\ell=1$, one gets
\begin{equation}
\label{eq:diffmixedreinforced}
\begin{aligned}
\de{W_\text{iso}}{I_1}\left(3,3\right)+2\de{W_\text{iso}}{I_2}\left(3,3\right)=0.
\end{aligned}
\end{equation}

Eqs.~\eqref{eq:difflreinforced}--\eqref{eq:diffmixedreinforced} represent some necessary
conditions for the passive energy density $W$ in order to produce the same
results with the two activation approaches on uniaxial deformations and on simple
shears. Eq.~\eqref{eq:difflreinforced} is notably severe: a particular
combination of the two partial derivatives of the energy has to be
constant for any $x$. For instance, in the case of fiber-reinforced Mooney-Rivlin
materials, where
\[
W(I_1,I_2,I_4)=c_1(I_1-3)+c_2(I_2-3)+f(I_4),
\]
it follows immediately from~\eqref{eq:difflreinforced} that
\[
c_1+\frac{c_2}{x}=x(c_1+c_2)\qRq \frac{c_2}{c_1+c_2}=-x\quad\text{for
  any $x\geq 1$},
\]
which is impossible. Hence, in the case of a fiber-reinforced Mooney-Rivlin
material active stress and active strain are never equivalent.

Also the important case where $W_\text{iso}$ depends only on
$I_1$ is always ruled out, since condition~\eqref{eq:diffmixedreinforced} becomes
$W'_\text{iso}(3)=0$
and by~\eqref{eq:difflreinforced} we get
\[
W'_\text{iso}(I_1)=0\quad\text{for any $I_1\geq 3$},
\]
whence $W_\text{iso}$ is a constant.

Coming back to the general fiber-reinforced case, we can also take the third derivative
of~\eqref{eq:Wlxreinforced}, twice w.r.t. $\ell$ and once
w.r.t. $x$. Letting $x=\ell=1$ we get the relation
\begin{multline}
\label{eq:thirdllx}
2\de{W_\text{iso}}{I_1}(3,3)+7\de{W_\text{iso}}{I_2}(3,3)
+8\dde{W_\text{iso}}{I_1^2}(3,3)\\
+12\dde{W_\text{iso}}{I_1\partial I_2}(3,3)
+4\dde{W_\text{iso}}{I_2^2}(3,3)=0.
\end{multline}
On the other hand, taking the third derivative 
twice w.r.t. $x$ and once
w.r.t. $\ell$ and letting $x=\ell=1$ we get
\begin{multline}
\label{eq:thirdlxx}
\de{W_\text{iso}}{I_1}(3,3)+3\de{W_\text{iso}}{I_2}(3,3)
+3\dde{W_\text{iso}}{I_1^2}(3,3)\\
+6\dde{W_\text{iso}}{I_1\partial I_2}(3,3)
+3\dde{W_\text{iso}}{I_2^2}(3,3)=0.
\end{multline}

By combining~\eqref{eq:diffmixedreinforced}, \eqref{eq:thirdllx}
and~\eqref{eq:thirdlxx} we get the following necessary condition involving the
second derivatives of the isotropic part of the energy density in the
identity deformation:
\begin{equation}
\label{eq:thirdcombined}
\dde{W_\text{iso}}{I_1^2}(3,3)
+6\dde{W_\text{iso}}{I_1\partial I_2}(3,3)
+5\dde{W_\text{iso}}{I_2^2}(3,3)=0.
\end{equation}
For instance, isotropic energy densities of the kind
\[
W_\text{iso}(I_1,I_2)=c_1(I_1-3)^2+c_{12}(I_1-3)(I_2-3)+c_2(I_2-3)^2
\]
satisfy the last condition only for a very particular choice of the elastic
moduli.

\section{A quantitative example}\label{sec:Fainco}

In this section we highlight that the differences between the two activation approaches 
can be considerable, even if the passive energy is quite simple, as in
the case of a Mooney-Rivlin material with a transversely isotropic
reinforcing term:
\begin{equation}
\label{eq:energy}
W_\text{pas}(\F)=c_1(I_1-3)+c_2(I_2-3)+c_3(\sqrt{I_4}-1)^2.
\end{equation}
Sect.~\ref{sec:energyebi} will be devoted to a more
refined energy which is commonly used for modeling skeletal
muscle tissue.

Let us assume an active strain of the form \eqref{eq:Fa} and deduce
the corresponding active part of the energy $W_\text{act}$ which gives
the same stress on the uniaxial deformation $\F_\lambda$~\eqref{eq:uniax}.
The energy density of the active strain approach is
\begin{multline*}
W_\text{strain}(\lambda;a)=W_\text{pas}(\F_\lambda\F_a^{-1})\\
=c_1\left(\frac{\lambda^2}{(1-a)^2}+\frac{2(1-a)}{\lambda}-3\right)\\
+c_2\left(\frac{(1-a)^2}{\lambda^2}+\frac{2\lambda}{1-a}-3\right)+
c_3\left(\frac{\lambda}{1-a}-1\right)^2,
\end{multline*}
and the passive part is given by
$W_\text{pas}(\lambda)=W_\text{strain}(\lambda;0)$. As in the previous
section, recalling that $W_\text{stress}=W_\text{pas}+W_\text{act}$,
the function $W_\text{act}$ such that the two energies coincide on the deformation
$\F_\lambda$ is
\begin{multline}
\label{eq:Wainc}
W_\text{act}(\lambda;a)=W_\text{strain}(\lambda;a)-W_\text{strain}(\lambda;0)\\
=a\left[c_1 \left(\frac{2-a}{(1-a)^2}\lambda^2-\frac{2}{\lambda}\right)
+c_2\left(\frac{2}{1-a}\lambda-\frac{2-a}{\lambda^2}\right)
+c_3\frac{\lambda}{1-a}\left(\frac{2-a}{1-a}\lambda-2\right)\right].
\end{multline}

\begin{figure}
\includegraphics[width=.49\textwidth]{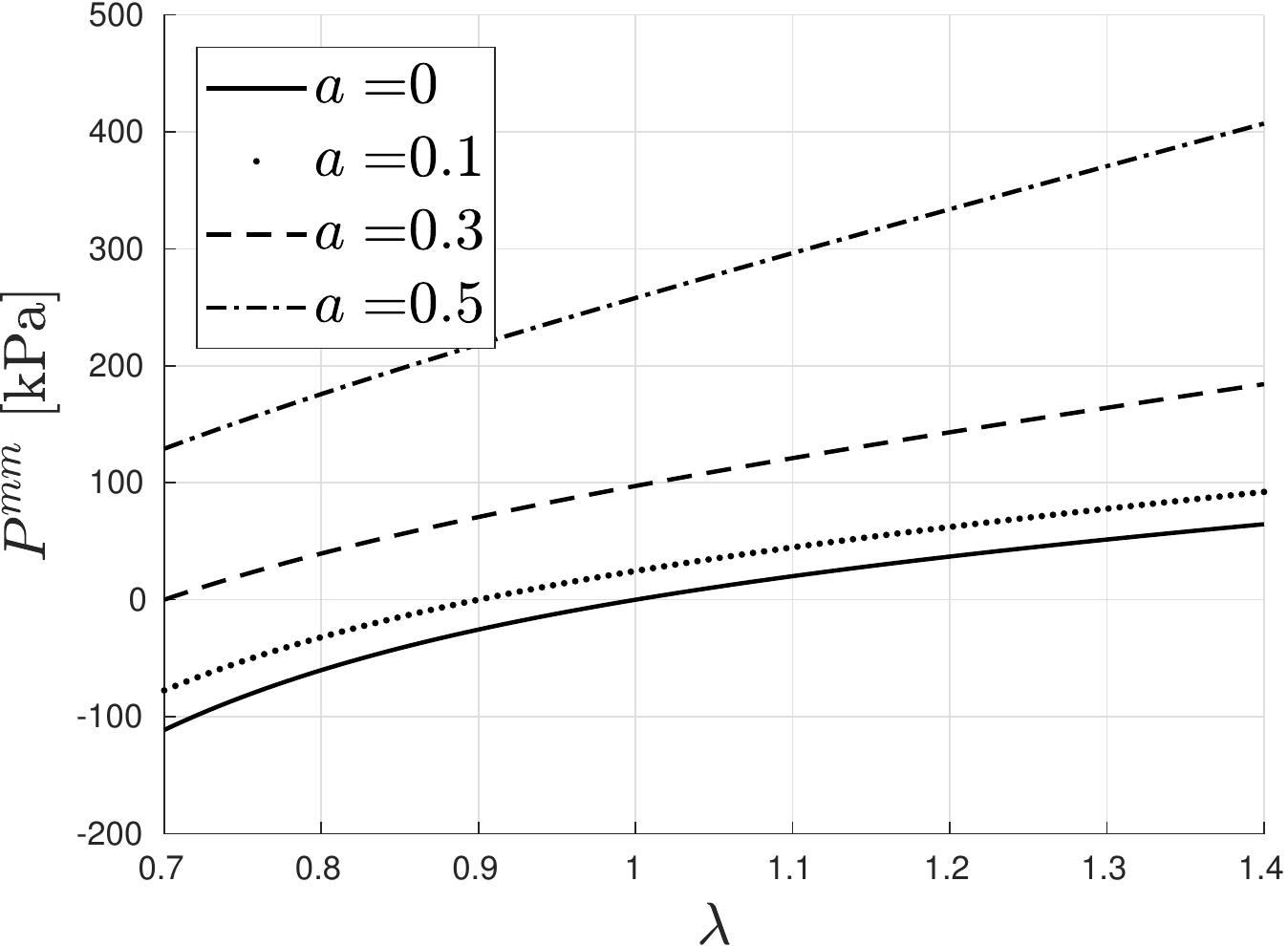}\hfill
\includegraphics[width=.49\textwidth]{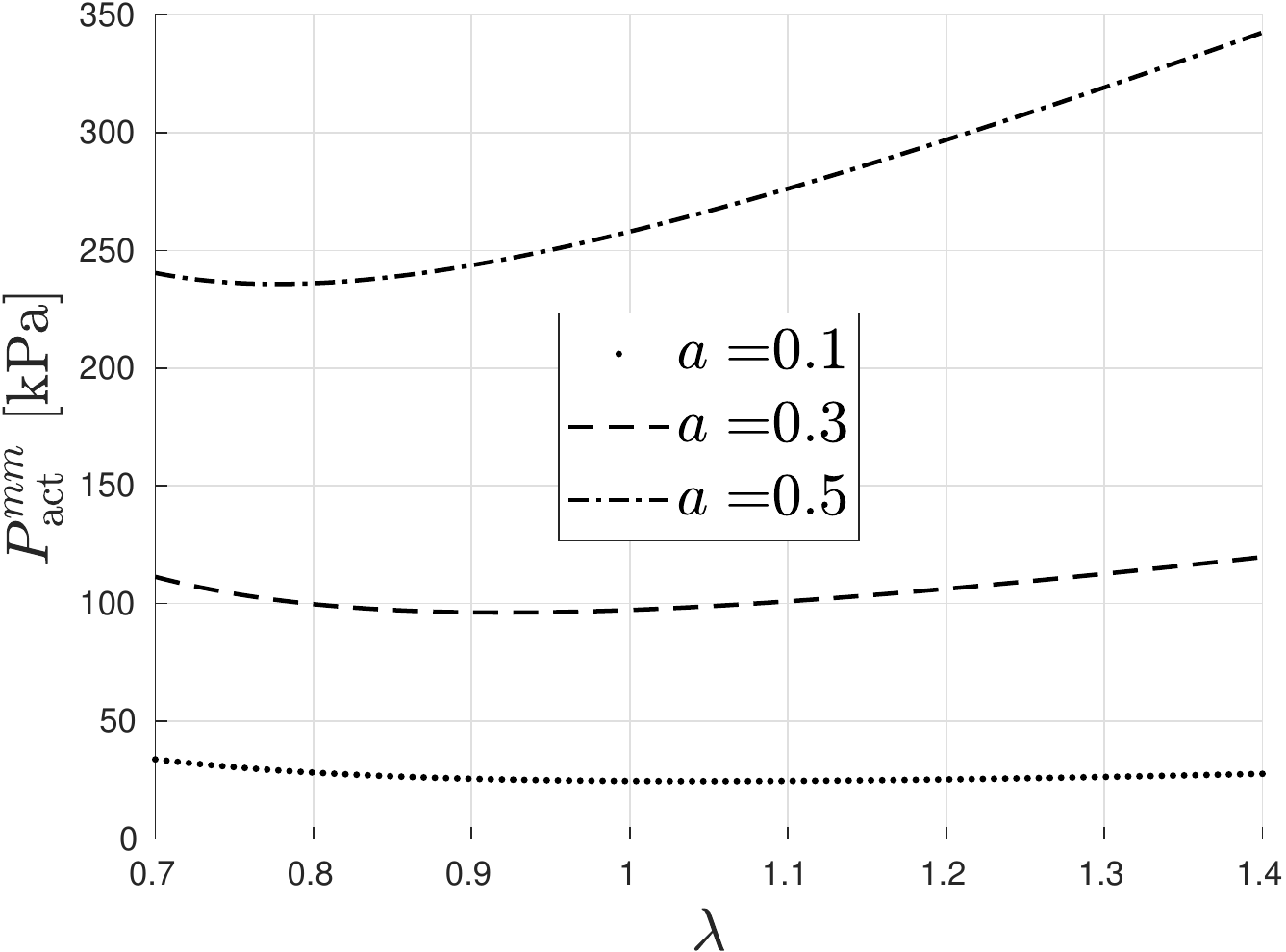}
\caption{On the left: stress-stretch relation obtained on the uniaxial
  deformation along the direction of anisotropy. On the right: active part of the stress.}
\label{fig:uniaxial}
\end{figure}

Fig.~\ref{fig:uniaxial} shows the profile of the stress and of its
active part along the uniaxial deformations for several values of the
activation parameter $a$. From now on we fix $c_1=4$kPa, $c_2=20$kPa, $c_3=40$kPa.
Denoting by $P^{mm}$ and $P_\text{act}^{mm}$ the components along
$\m\otimes\m$ of the total and active stress, respectively, one can
see that $P^{mm}$ and $P_\text{act}^{mm}$ increase, as the
parameter $a$ increases.

On a general deformation $\F$, the stresses corresponding to the active
strain and the active stress models will be directly computed
using~\eqref{eq:astrain} and~\eqref{eq:astress}, respectively.
Taking into account incompressibility, we have
{\footnotesize
  \begin{align*}
\P_\text{strain}=&\left(2c_1\F_e+2c_2(I_1(\C_e)\F_e-\F_e\F_e^\T\F_e)+2c_3(\sqrt{I_4(\C_e)}-1)\frac{1}{\sqrt{I_4(\C_e)}}\F_e\m\otimes\m\right)\F_a^{-\T}\\
&-p_\text{strain}\F^{-\T},\\
\P_\text{stress}=&2c_1\F+2c_2(I_1\F-\F\F^\T\F)+2c_3(\sqrt{I_4}-1)\frac{1}{\sqrt{I_4}}\F\m\otimes\m
+\frac{W_\text{act}'(\sqrt{I_4})}{\sqrt{I_4}}\F\m\otimes\m\\
&-p_\text{stress}\F^{-\T}.
\end{align*}
}

Let us analyze the response of the two approaches on the simple shear
$\F_K$ given by \eqref{eq:shear}.  We will follow the classical
assumption of plane stress in order to find the unknown pressure
fields $p_\text{strain}$ and $p_\text{stress}$, that is
$P_\text{strain}^{ss}=0$ and $P_\text{stress}^{ss}=0$, where
$\s=\m\times\n$. Another possibility, which will not be taken into
consideration in this paper, is to assume zero normal traction on the
inclined faces; for a discussion, see \cite{Horgan2011}.

\begin{figure}
\includegraphics[width=.49\textwidth]{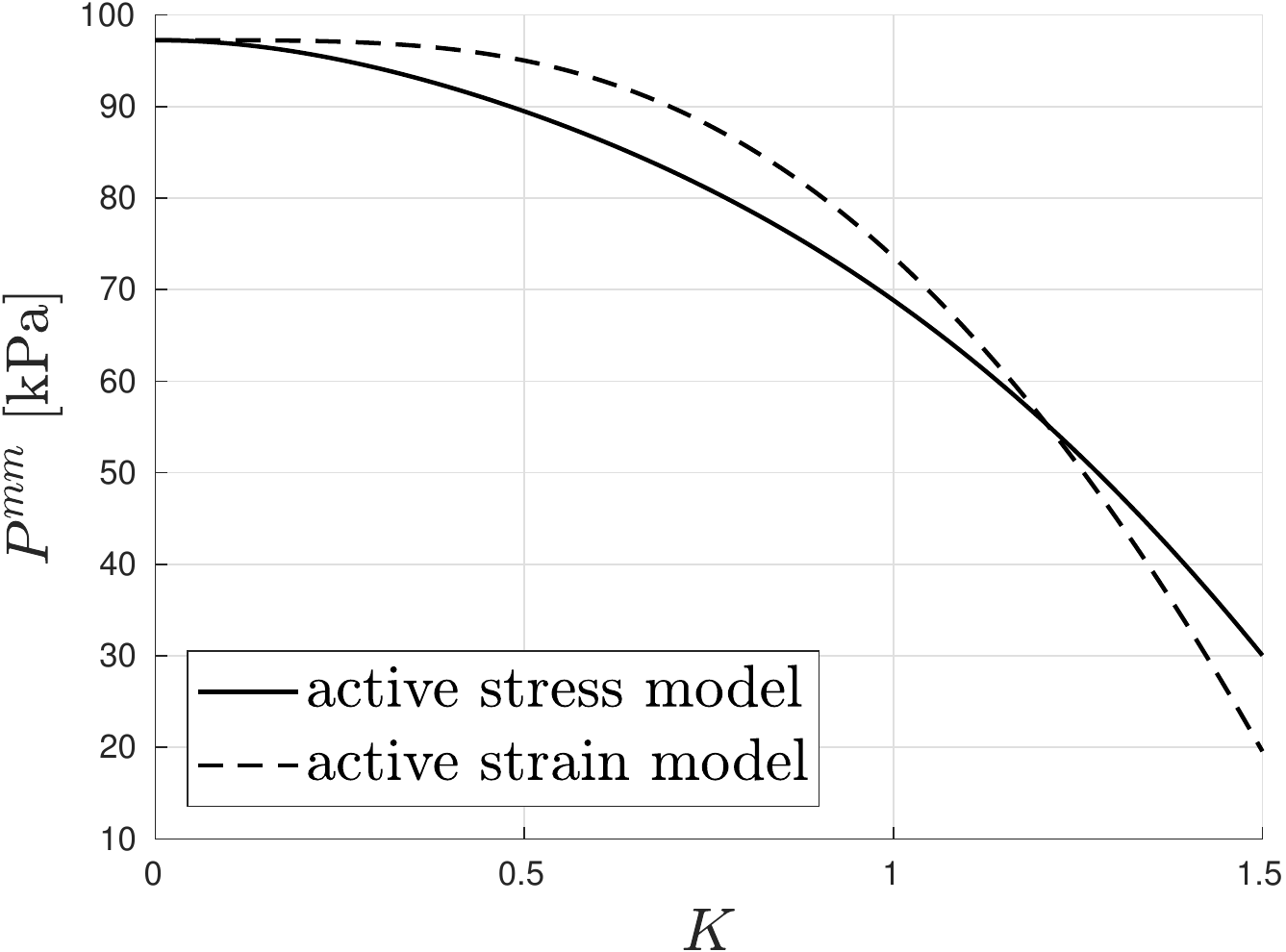}
\includegraphics[width=.49\textwidth]{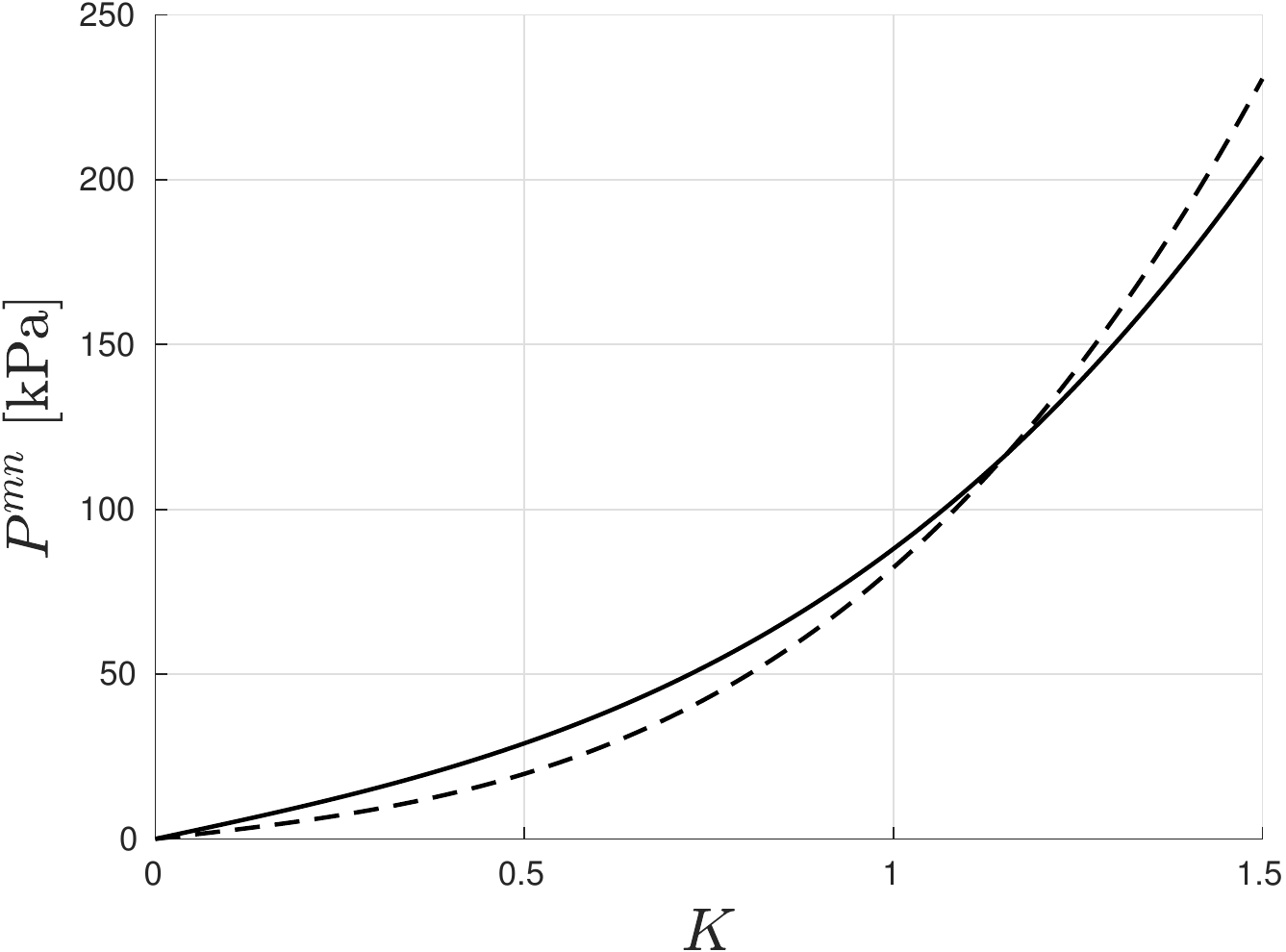}\\
\includegraphics[width=.49\textwidth]{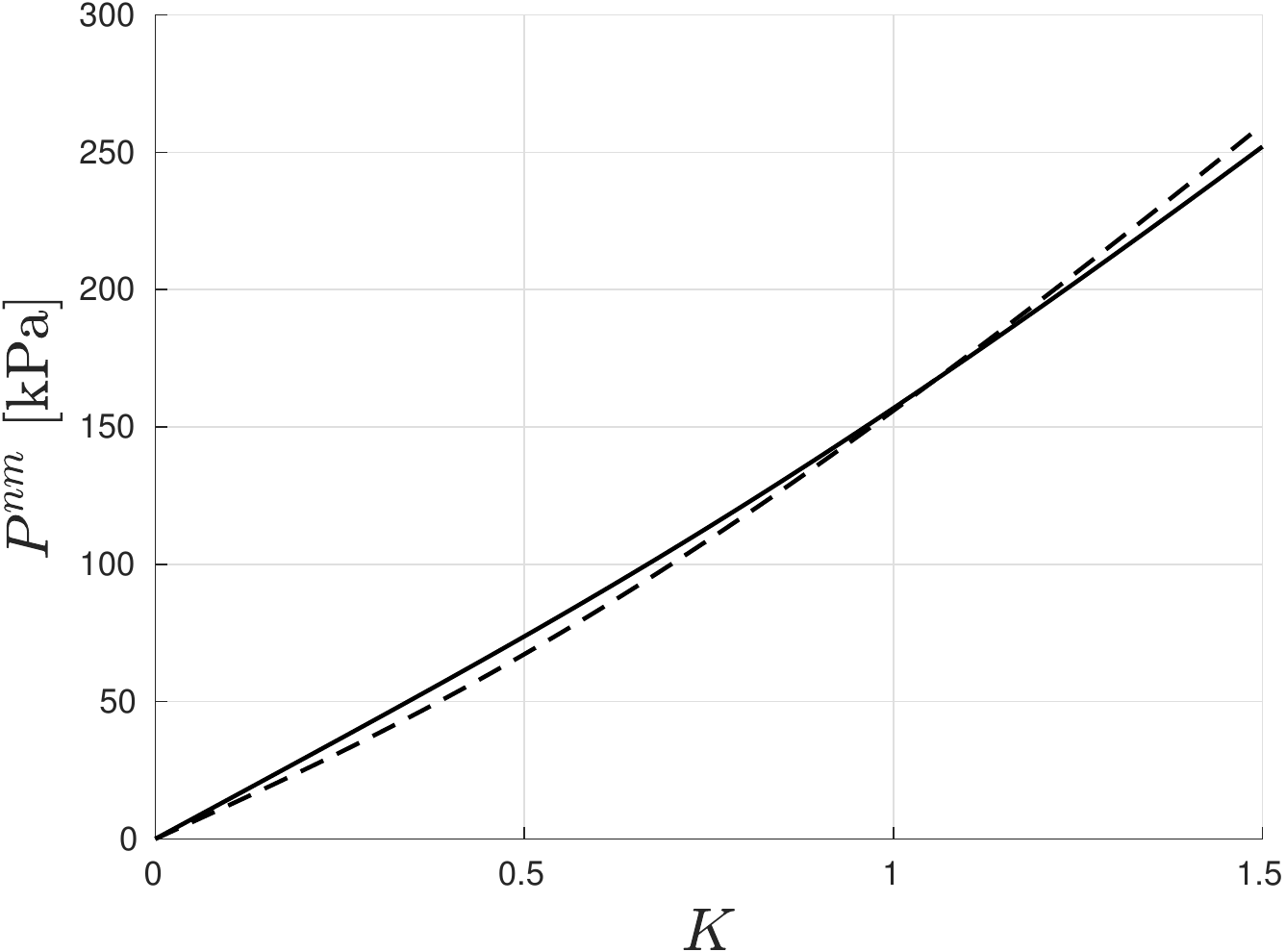}
\includegraphics[width=.49\textwidth]{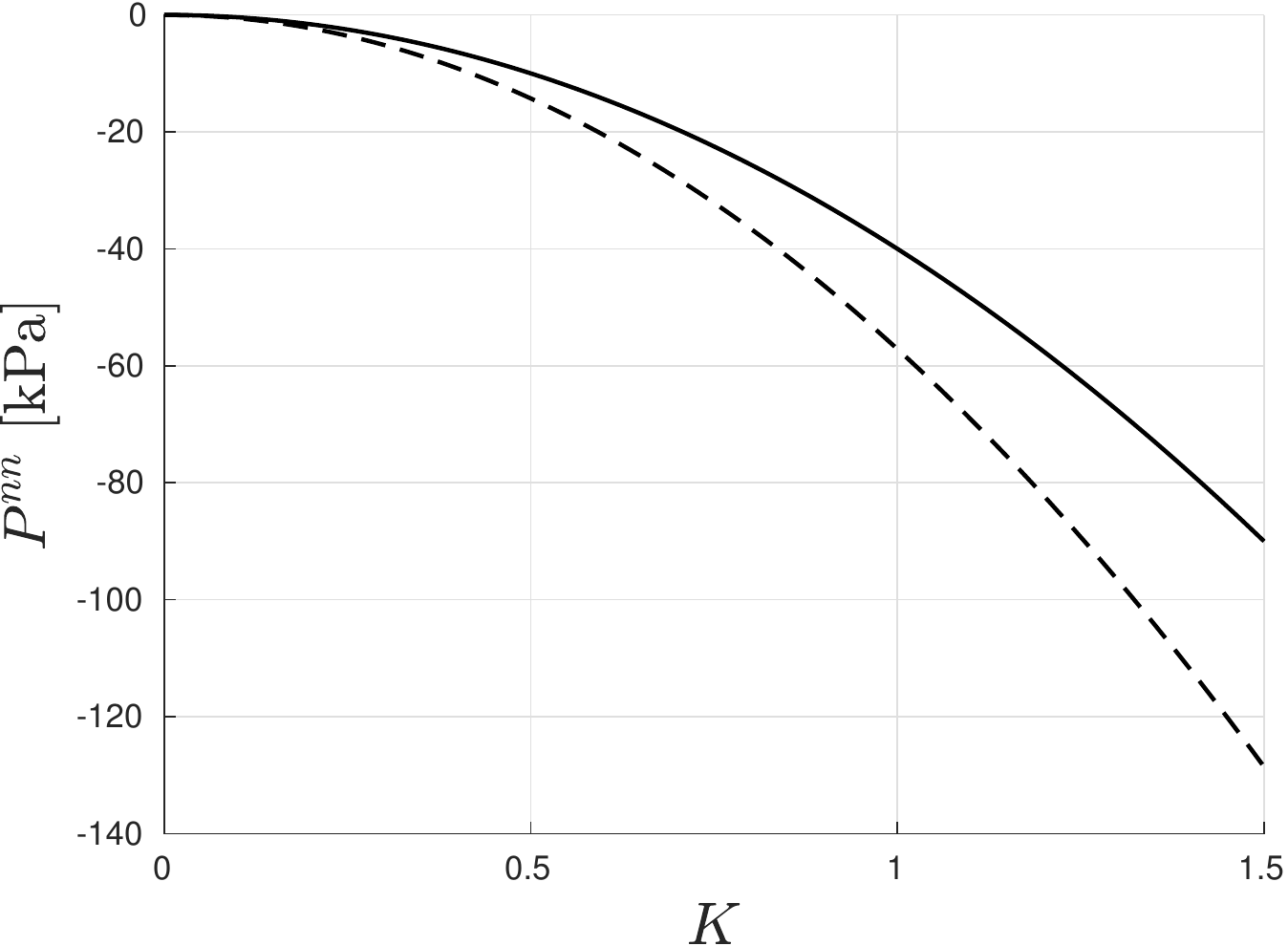}
\caption{Comparison between the stress components of the two
activation approaches when $a=0.3$.}
\label{fig:WincoFaincocomp}
\end{figure}

\begin{figure}
\includegraphics[width=.49\textwidth]{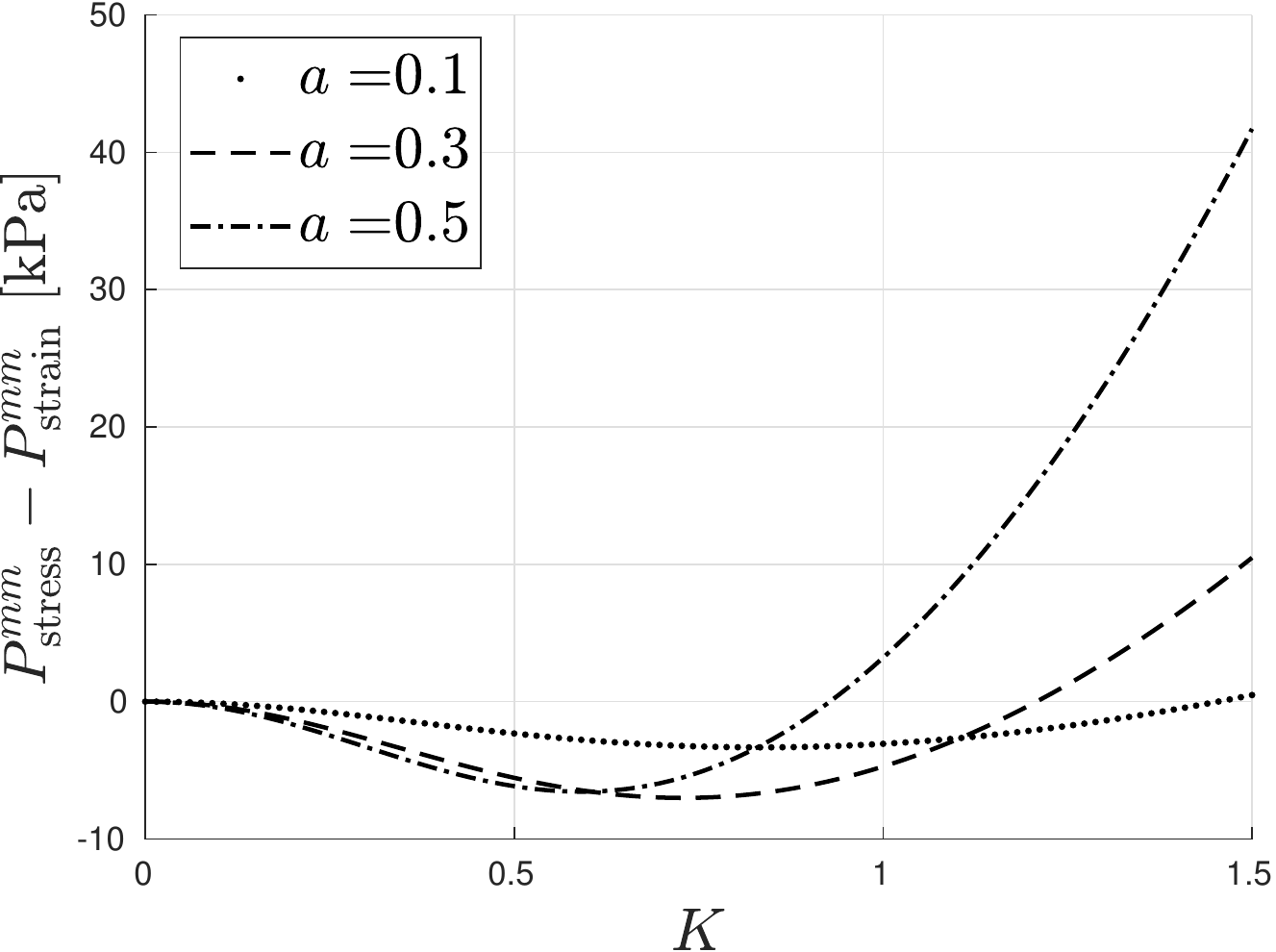}
\includegraphics[width=.49\textwidth]{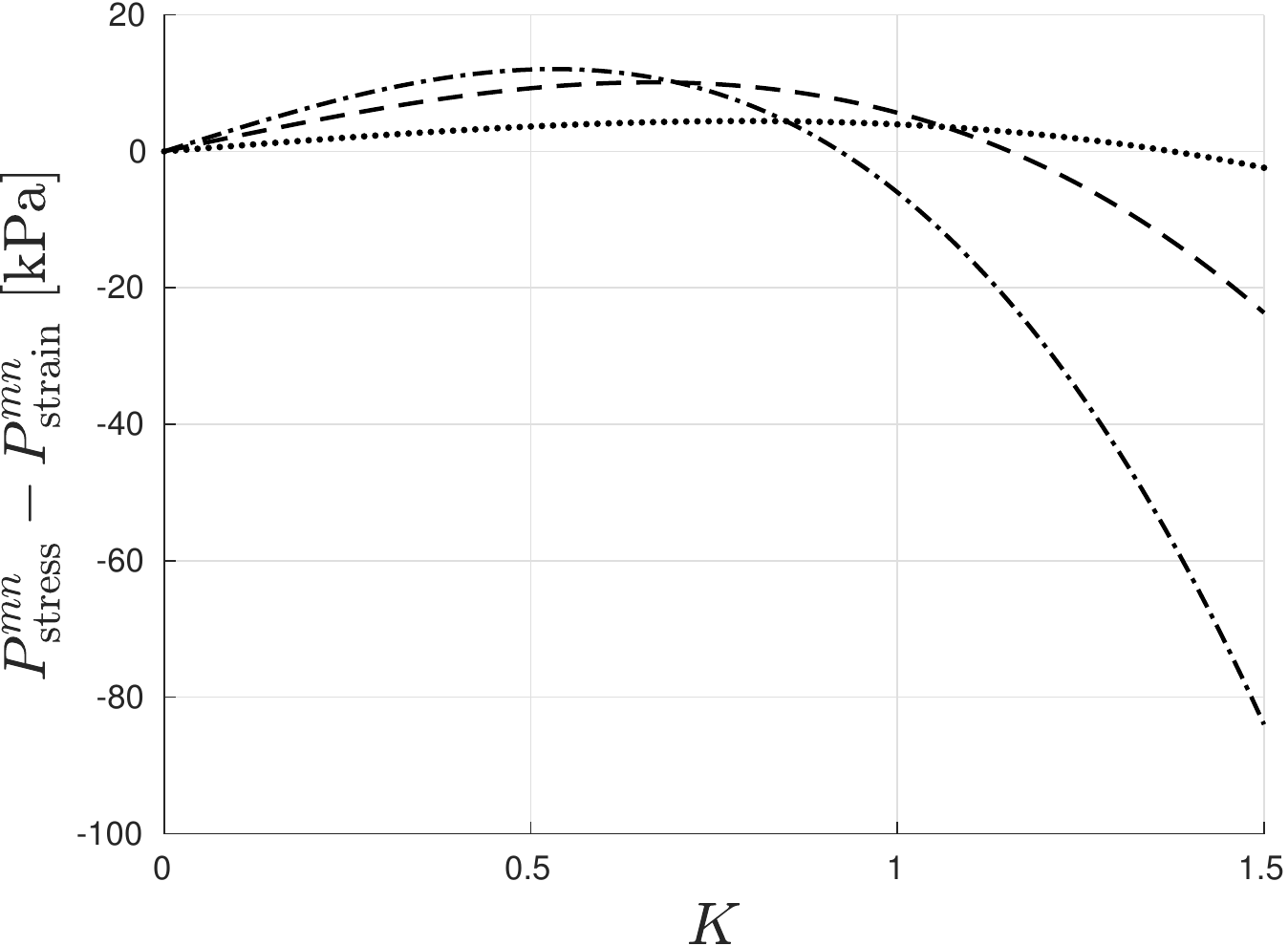}\\
\includegraphics[width=.49\textwidth]{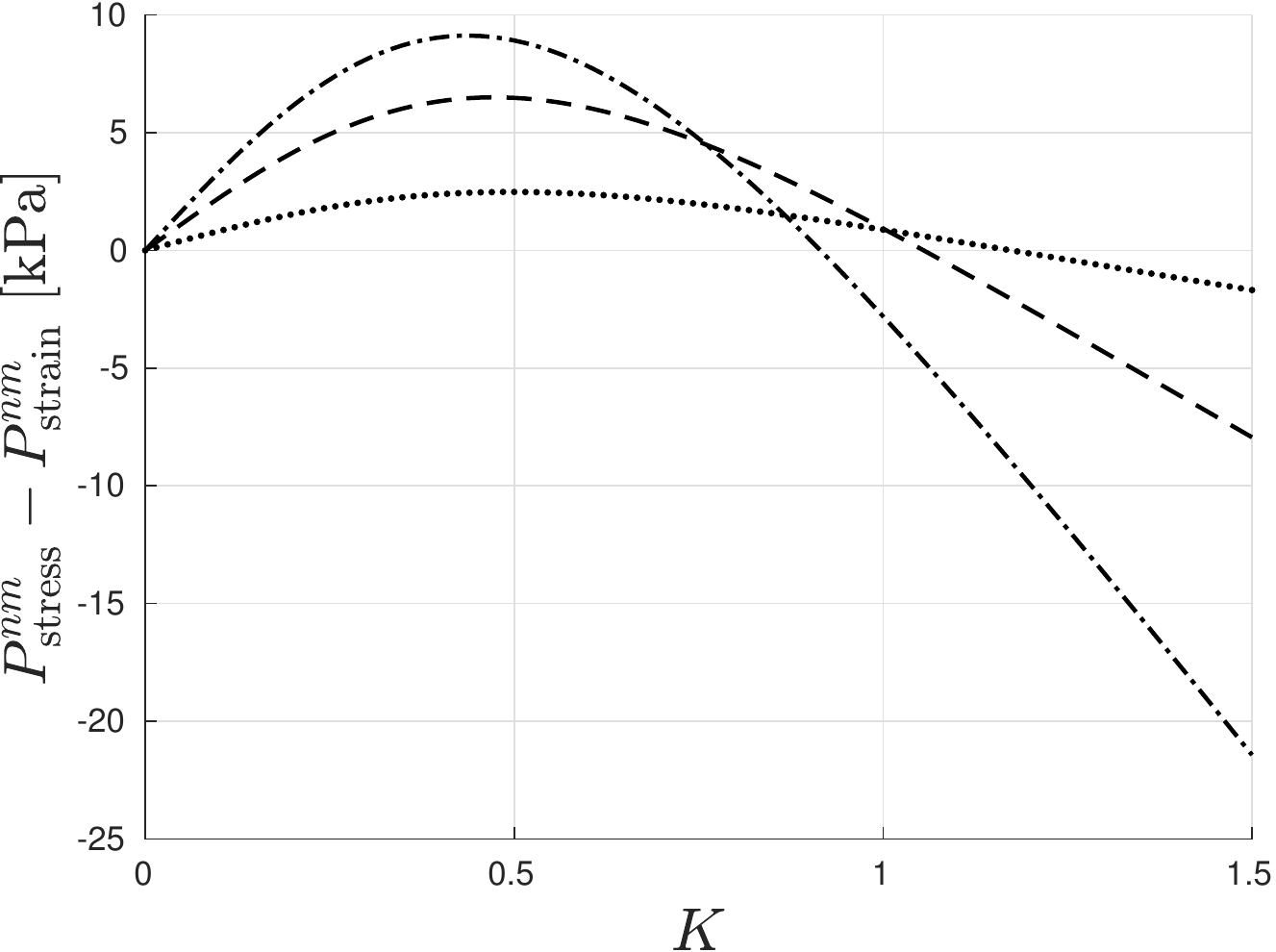}
\includegraphics[width=.49\textwidth]{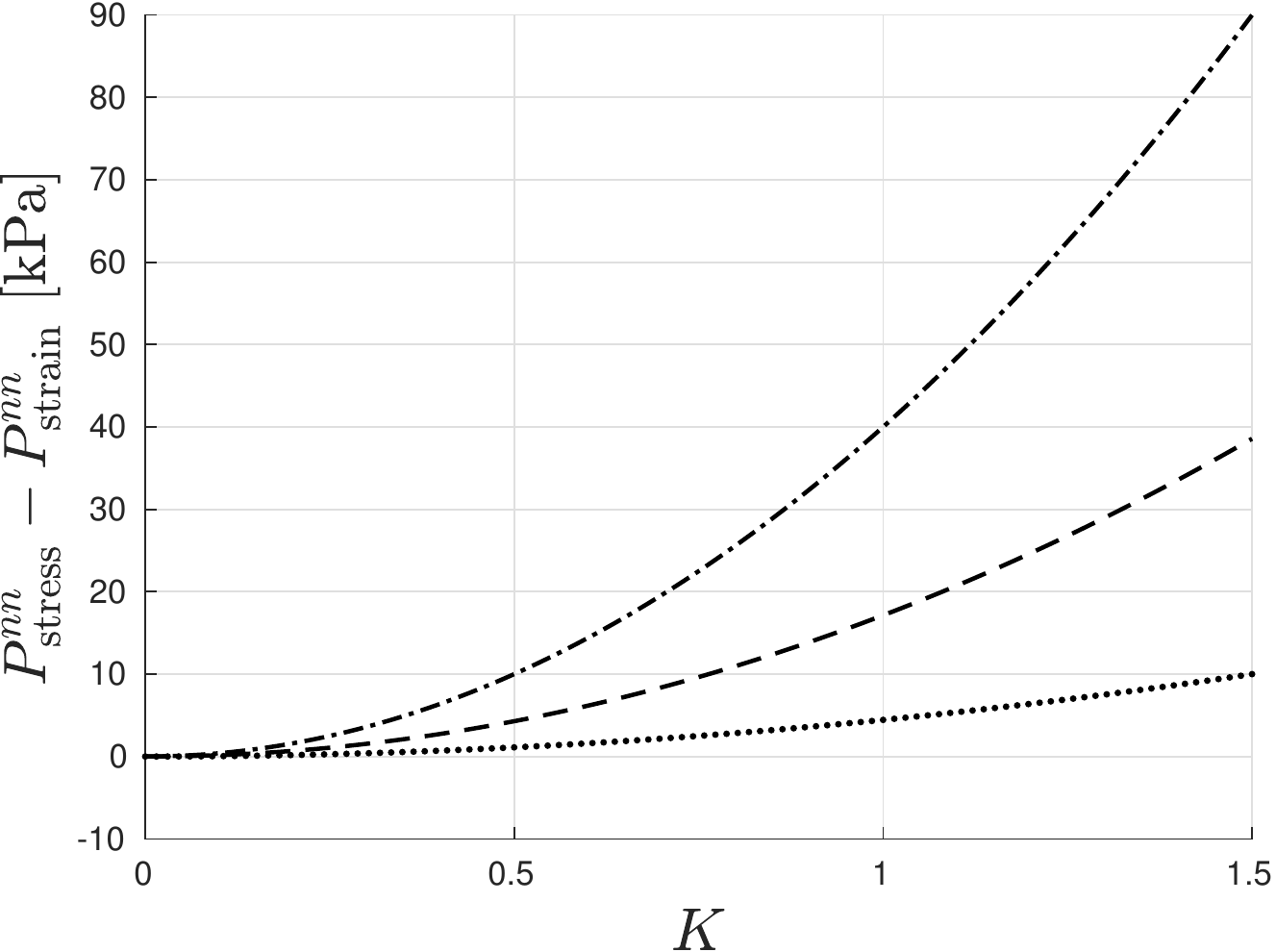}
\caption{Differences between the components of the stress in the two
 activation approaches for several value of the activation parameter $a$.}
\label{fig:WincoFaincodifferenze}
\end{figure}

The non-vanishing
components of the stresses are given by
\begin{footnotesize}
\begin{equation}\label{eq:compK}
\begin{aligned}
P_\text{strain}^{mm}=&2c_1\left(\frac{1}{(1-a)^2}-(1-a)\right)
+2c_2\left(\frac{1-K^2}{1-a}-(1-a)^2\right)+\frac{2c_3}{1-a}\left(\frac{1}{1-a}-\frac{1}{\sqrt{1+K^2}}\right),\\[1ex]
P_\text{stress}^{mm}=&2c_1a\left(\frac{2-a}{(1-a)^2}+\frac{1}{(1+K^2)^{3/2}}\right)
+2c_2\left(\frac{a(2-a)}{(1+K^2)^2}+\frac{a}{(1-a)\sqrt{1+K^2}}-K^2\right)\\[1ex]
&+\frac{2c_3}{1-a}\left(\frac{1}{1-a}-\frac{1}{\sqrt{1+K^2}}\right),\\[1ex]
P_\text{strain}^{mn}=&2c_1K(1-a)+2c_2 K\left(\frac{K^2}{1-a}+(1-a)^2\right),\\[1ex]
P_\text{stress}^{mn}=&2K\Big(c_1+c_2(1+K^2)\Big)=P_\text{pas}^{mn},\\[1ex]
P_\text{strain}^{nm}=&2c_1\frac{K}{(1-a)^2}+2c_2\frac{K}{1-a}+2c_3\frac{K}{(1-a)}\left(\frac{1}{1-a}-\frac{1}{\sqrt{1+K^2}}\right),\\[1ex]
P_\text{stress}^{nm}=&2c_1K\left(\frac{1}{(1-a)^2}+\frac{a}{(1+K^2)^{3/2}}\right)
+2c_2 K\left(1+\frac{a(2-a)}{(1+K^2)^2}+\frac{a}{(1-a)\sqrt{1+K^2}}\right)\\[1ex]
&+2c_3\frac{K}{1-a}\left(\frac{1}{1-a}-\frac{1}{\sqrt{1+K^2}}\right),\\[1ex]
P_\text{strain}^{nn}=&-2c_2\frac{K^2}{1-a},\\[1ex]
P_\text{stress}^{nn}=&-2c_2 K^2= P_\text{pas}^{nn}.
\end{aligned} 
\end{equation}
\end{footnotesize}
In Fig.~\ref{fig:WincoFaincocomp} we plot such components with respect
to the amount of shear $K$, both in the case of active strain and of
active stress. As one can see, even if the two activation approaches
produce the same stress tensor on uniaxial deformations along the
direction of anisotropy, the stresses are different on the simple shear
$\F_K$.
The dependence of the differences between $\P_\text{stress}$ and
$\P_\text{strain}$ on the activation parameter $a$ is showed in
Fig.~\ref{fig:WincoFaincodifferenze}: such a difference is more
evident when $a$ increases. As we have already noticed, the active
part of $\P_\text{stress}$ lies along $\F_K\m\otimes\m$, so that
$P^{mn}_\text{stress}$ and $P^{nn}_\text{stress}$ in \eqref{eq:compK}
do not depend on $a$. Hence the plots on the right in
Fig.~\ref{fig:WincoFaincodifferenze} represent the differences between
$\P_\text{pas}$ and $\P_\text{strain}$ along $\m\otimes\n$ and
$\n\otimes\n$.

\section{A more complex energy related to skeletal muscle tissue}
\label{sec:energyebi}

As an important example, we now consider the case of the activation of
a skeletal muscle tissue, for which there are several experimental
data on uniaxial deformations. In this case it is easier to measure
the active stress $\P_\text{act}$ than the active strain $\F_a$:
indeed, the components of the active stress can be obtained by
computing the difference between the data collected in the active and
passive case, see for instance~\cite{wilkie56,hawkins}. Hence,
differently from the previous sections, we start from a
given active stress and find a suitable active strain that
produces the same results in the uniaxial deformations along the
direction of anisotropy.

As far as the passive energy and the active stress are concerned, we
will follow the model given in~\cite{ebi}, while the active strain
will be modeled as in~\cite{gammaderiv}. We will compare the
stress components along a cross-fiber simple shear obtained by
exploiting the two activation approaches.

\begin{figure}
        \includegraphics[width=0.70\textwidth]{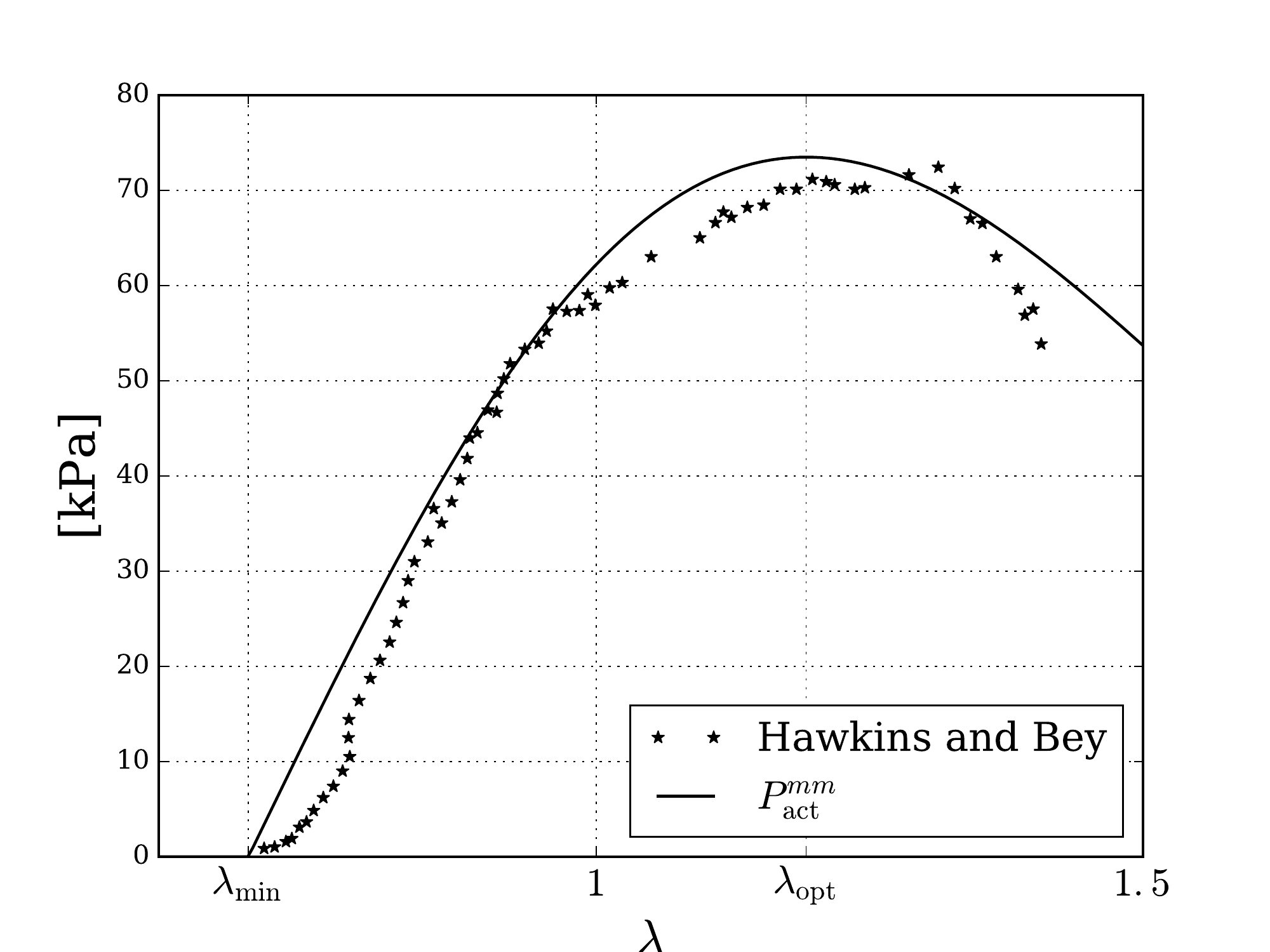}
   \caption{Plot of~\eqref{eq:pact} together with the representation 
   of the experimental data given in \cite{hawkins}.} 
     \label{fig:attiva}
\end{figure}

The typical active stress-stretch curve
reaches a maximum point at $\lambda_{\text{opt}}$ and then decreases
for larger values of $\lambda$, see for instance \cite{hawkins} for
the tetanized \emph{tibialis anterior} of a rat.
Following~\cite{ebi}, we
assume that the active part of the stress is
\begin{equation}\label{eq:pact}
 P_\text{act}^{mm}(\lambda) =
\left\{
\begin{aligned}
&P_{\text{opt}}\frac{\lambda_{\text{min}}-\lambda}{\lambda_{\text{min}}-\lambda_{\text{opt}}}e^{\frac{(2\lambda_{\text{min}}-\lambda-\lambda_{\text{opt}})(\lambda-\lambda_{\text{opt}})}{2(\lambda_{\text{min}}-\lambda_{\text{opt}})^2}}
& \text{if $\lambda>\lambda_{\text{min}}$},
\\[1ex]
&0 & \text{otherwise},
\end{aligned}
\right. 
\end{equation}
where $\lambda_{\text{min}}$ is the minimum stretch value after which
the activation starts, while $(\lambda_{\text{opt}},P_{\text{opt}})$
identifies the maximum of the curve. According to \cite{ebi}, we set
$\lambda_{\text{min}}=0.682$, $\lambda_{\text{opt}}=1.192$,
$P_{\text{opt}}=73.52$kPa.  Fig.~\ref{fig:attiva} shows that the curve
fits quite well the active data obtained in \cite{hawkins}.

Since~\eqref{eq:pact} represents the component of
$\P_\text{act}$ along $\F\m\otimes\m$, for the active stress case it is
enough to compute a primitive function in order to write the energy
density $W_\text{act}$. Denoting by $\sqrt{I_4}$ the stretch along
the direction of anisotropy $\m$ in a general deformation, we have
\begin{small}
\[
 W_\text{act}(\sqrt{I_4})=
\left\{
\begin{aligned}
&P_{\text{opt}}(\lambda_{\text{min}}-\lambda_{\text{opt}})
\left(e^{\frac{(2\lambda_{\text{min}}-\sqrt{I_4}-\lambda_{\text{opt}})(\sqrt{I_4}-\lambda_{\text{opt}})}
    {2(\lambda_{\text{min}}-\lambda_{\text{opt}})^2}}-e^{1/2}\right)
& \text{if $\lambda>\lambda_{\text{min}}$},\\
&0 & \text{otherwise}.
\end{aligned}
\right. 
\]
\end{small}
Notice that the experimental
data show that the active part of the stress is not monotone, hence
the corresponding total strain energy can lose the rank-one convexity.

\null

As far as the passive part is concerned, following again the model
given in~\cite{ebi} and~\cite{gammaderiv}, we use the exponential
strain energy density function
\begin{equation}
W_\text{pas}^\text{muscle}=\frac{\mu}{4}\left\{\frac{1}{\alpha}\left[e^{\alpha({I}_p-1)}-1\right]+
\frac{1}{\beta}\left[e^{\beta(K_p-1)}-1\right]\right\},\label{eq:Webi}
\end{equation}
where 
\begin{equation*}
  \begin{aligned}
{I}_p&=\frac{w_0}{3}\tr({\C})+(1-w_0)\tr({\C}{\m\otimes\m}), \\
K_p&=\frac{w_0}{3}\tr({{\C}^{-1}})+(1-w_0)\tr({\C}^{-1}{\m\otimes\m})
\end{aligned}
\end{equation*}
(here $\m$ is the direction of the muscular fibers).
Notice that in the incompressible case $I_p$ and $K_p$ can be
expressed in terms of the usual invariants as
\begin{equation*}
{I}_p=\frac{w_0}{3}I_1+(1-w_0)I_4, \ \
K_p=\frac{w_0}{3}I_2+(1-w_0)(I_5-I_1I_4+I_2). 
\end{equation*}
The material parameters $\alpha=19.69$, $\beta=1.190$, $w_0=0.7388$
and $\mu=0.1599$kPa given in \cite{ebi} are obtained from the passive
data about the \emph{tibialis anterior} of a rat \cite{hawkins}; in
particular, $w_0$ measures the amount of anisotropy of the material.

Now that the active stress and the passive energy have been chosen, we
want to find a suitable active strain which gives the same results on
uniaxial deformations. The issue is subtle because we cannot assume
that $\F_a$ is constant. Indeed one can see that a constant active
strain cannot fit at all the experimental data.
To address this problem we will assume that $\F_a$ depends
on the deformation gradient $\F$, see also
\cite{ebi, noi, gammaderiv, papersimai, WIMJ14}. Such an approach,
which is a generalization of the active strain, is crucial in the
applications to skeletal muscle: it allows to capture
the physics of a muscle, in which the stress produced when the tissue
is activated depends on strain.
In this case the mathematical properties of $W_\text{strain}$ can 
change considerably and $\F_a$ does not represent anymore the local distortion of the material that maps the reference configuration to the relaxed one. Moreover, the expression of the stress tensor is much more involved,
see~\cite{gammaderiv}:
\begin{equation}\label{eq:Pstrainnoncost}
P_\text{strain}^{hk}(\F)=\de{W_\text{strain}}{F^{hk}}(\F;\F_a(\F))+
\de{W_\text{strain}}{F_a^{ij}}(\F;\F_a(\F))
\de{F_a^{ij}}{F^{hk}}(\F)-p_\text{strain}(\F^{-\T})^{hk},
\end{equation}
with $h,k=1, 2, 3$.

\begin{figure}
\centering
\includegraphics[width=0.6\textwidth]{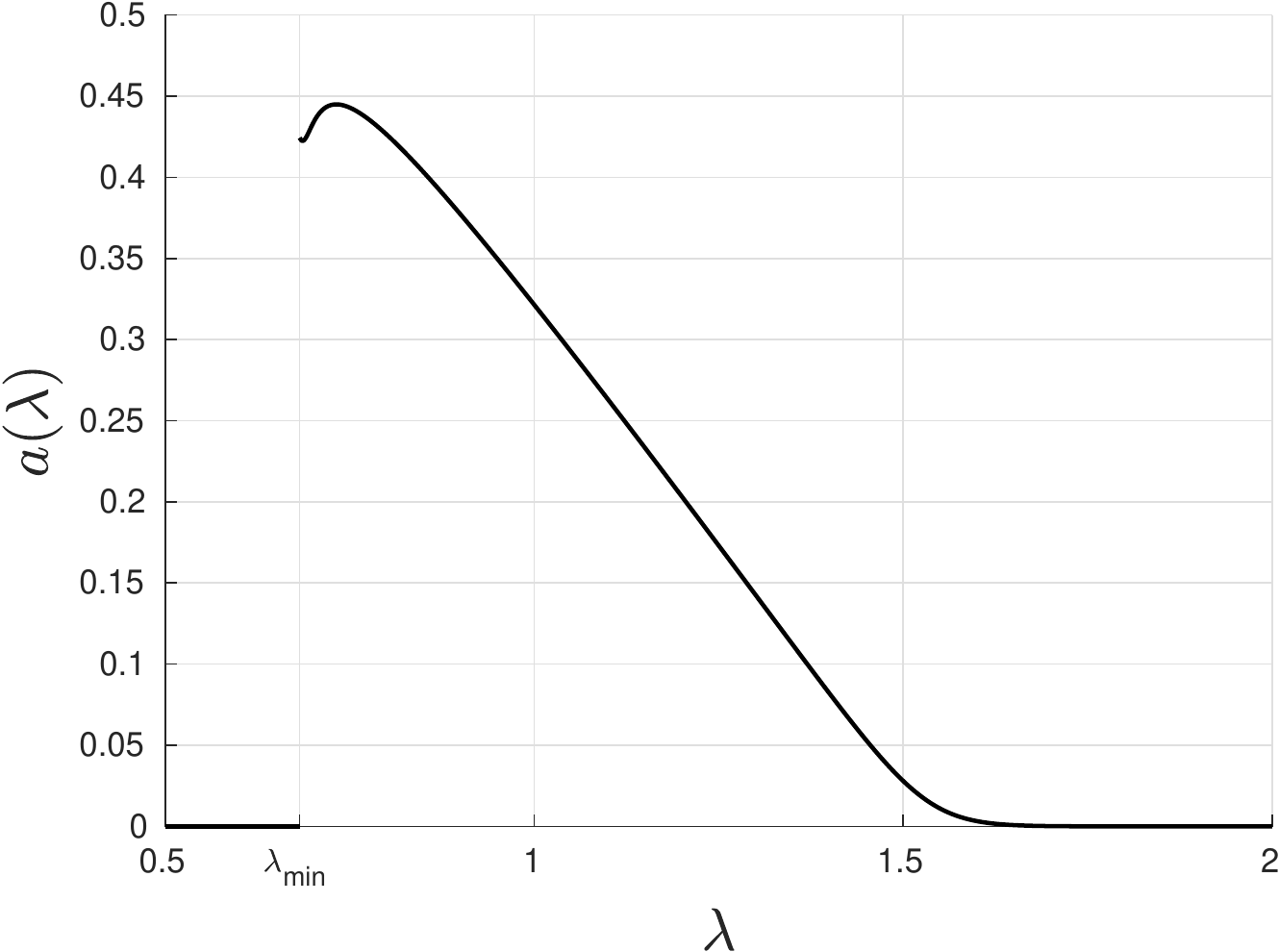}
\caption{Plot of the function $a(\lambda)$ solution of~\eqref{eq:daStressaStrain}.}
\label{fig:a_att_inc}
\end{figure}

Given an active strain $\F_a$ as in~\eqref{eq:Fa}, we  assume that
the activation parameter $a$ depends on the stretch of the
fibers. Along the uniaxial deformation
$\F_\lambda$~\eqref{eq:uniax} we have that $a$ is a function of $\lambda$.
Imposing that the energies of the active stress and of
the active strain formulation coincide on $\F_\lambda$, we look for
$a(\lambda)$ such that
\begin{equation}
\label{eq:daStressaStrain}
W_\text{strain}(\lambda;a(\lambda))=W_\text{pas}^\text{muscle}(\lambda)+
W_\text{act}(\lambda).
\end{equation}
Since the parameter $a$ accounts for a contraction, we recall that $0\leq
a(\lambda)<1$. Eq.~\eqref{eq:daStressaStrain} admits the solution
$a(\lambda)=0$ whenever $W_\text{act}(\lambda)=0$, but in general it is too
complicated to solve analytically. In Fig.~\ref{fig:a_att_inc} we plot
a numerical approximation of the solution $a(\lambda)$: the function is discontinuous
at $\lambda_\text{min}$ and vanishes as $\lambda\to+\infty$.

\begin{figure}
\includegraphics[width=.49\textwidth]{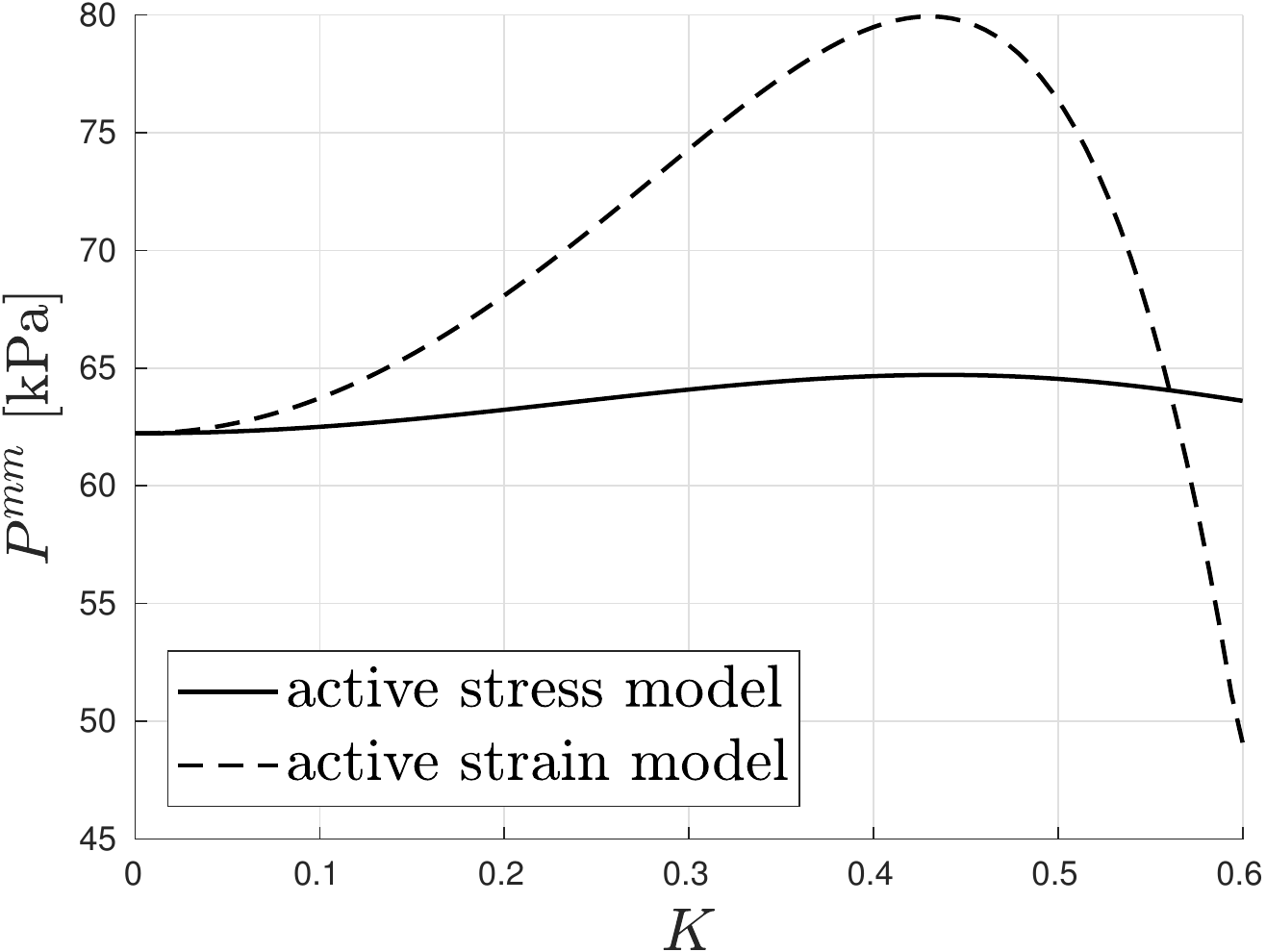}
\includegraphics[width=.49\textwidth]{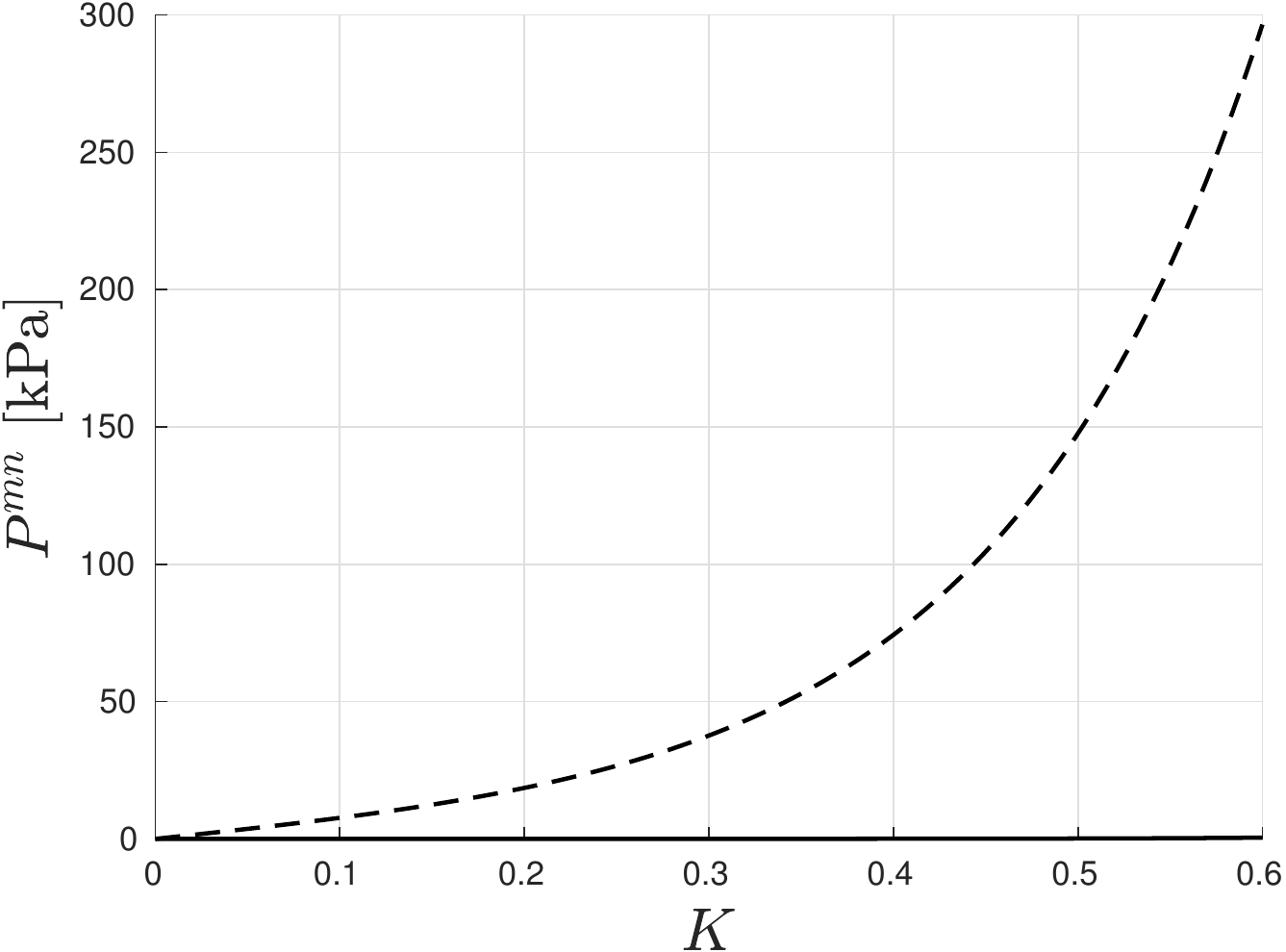}\\
\includegraphics[width=.49\textwidth]{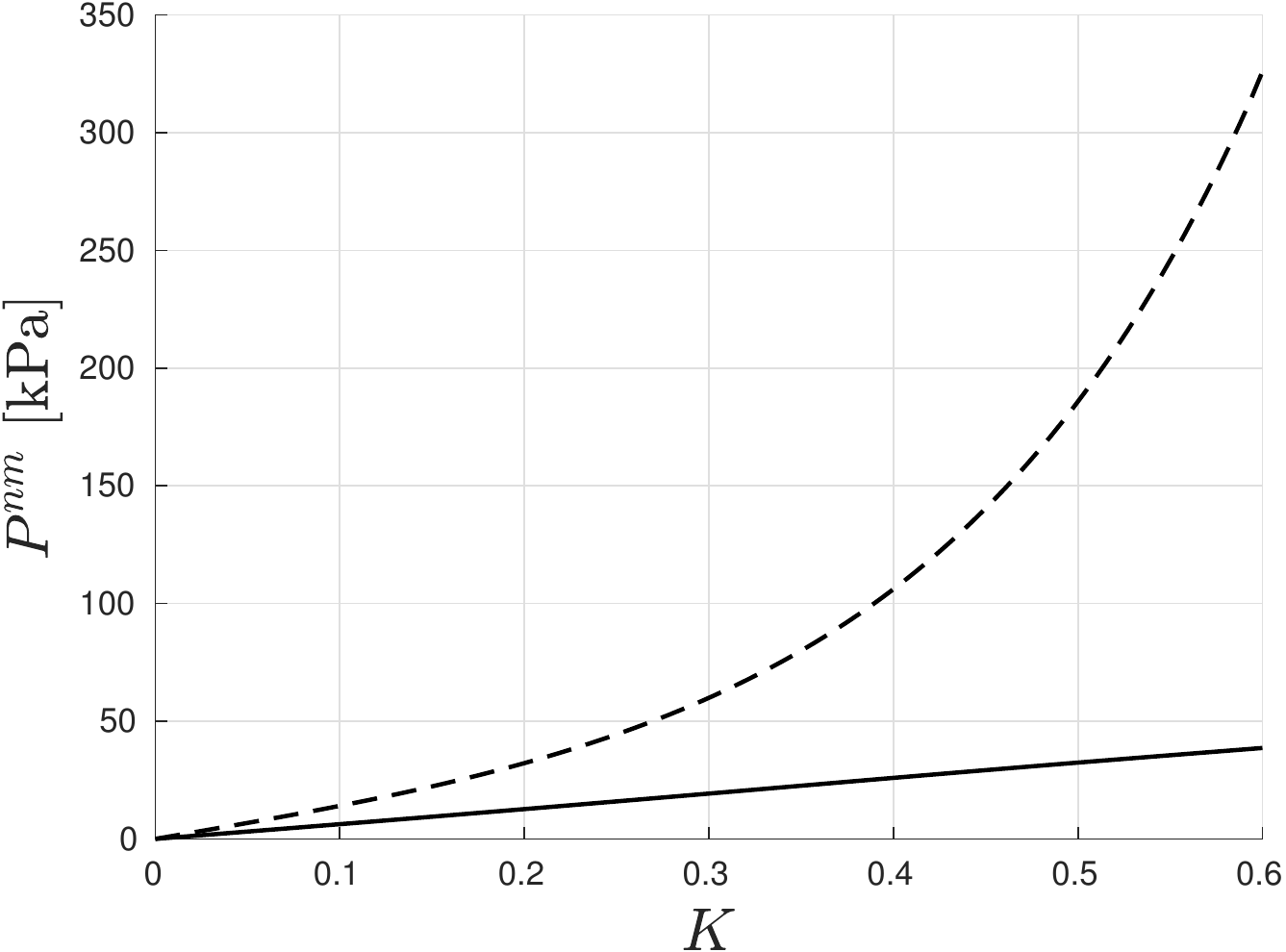}
\includegraphics[width=.49\textwidth]{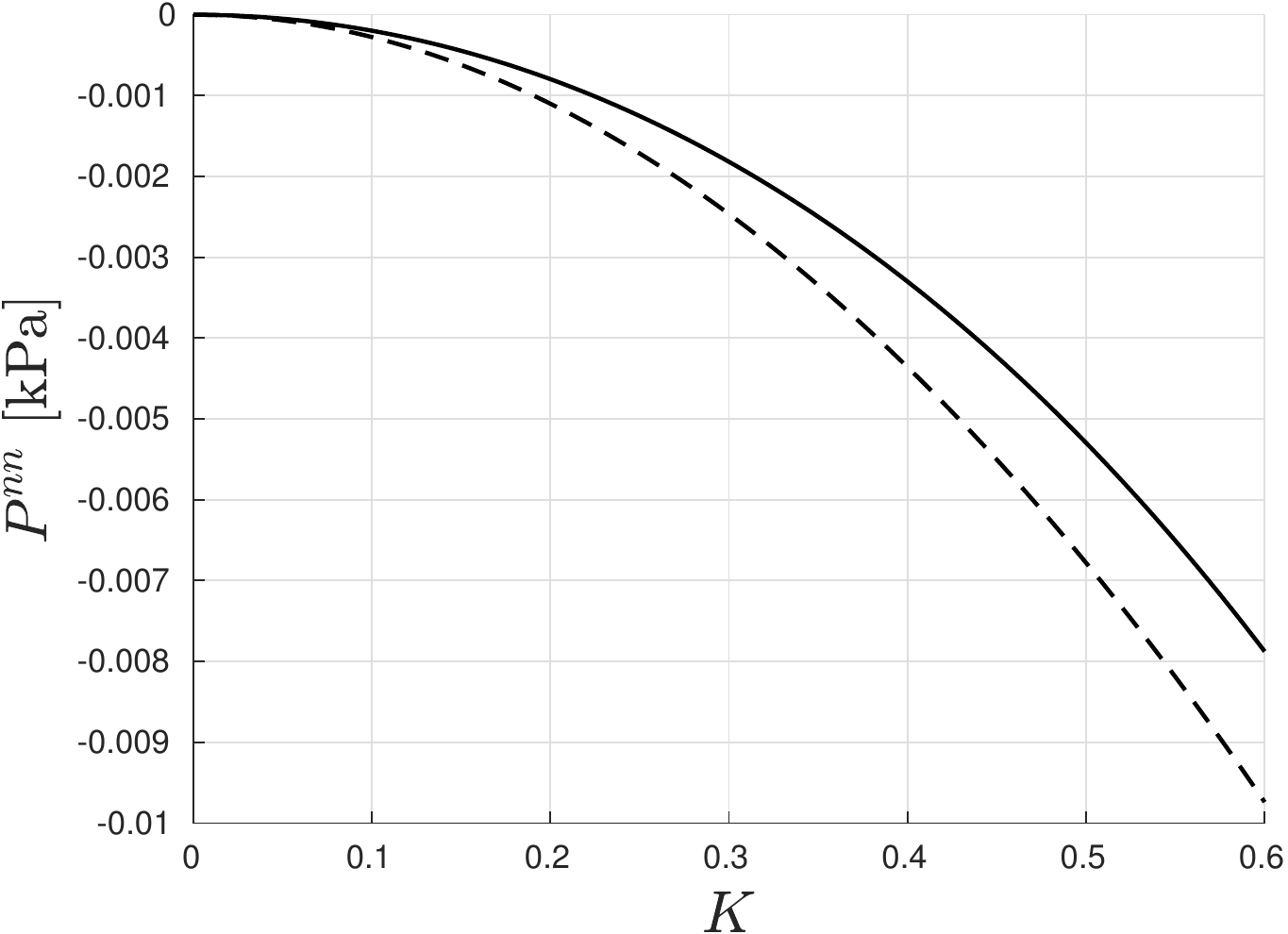}
\caption{Comparison between the stress components of the two
activation approaches in the case of energy \eqref{eq:Webi},
Sect.~\ref{sec:energyebi}. Here we considered only values of $K$ up to
$0.6$, which is the physiological range for a skeletal muscle tissue.} 
\label{fig:stressnostra}
\end{figure}

Now that the two approaches give the same stress along the uniaxial
deformations $\F_\lambda$, let us consider the simple shear
$\F_K$~\eqref{eq:shear}. Assuming as in the previous section that
$P_\text{strain}^{ss}=0$ and $P_\text{stress}^{ss}=0$, where
$\s=\m\times\n$, we can find the Lagrange multipliers related to the
incompressibility constraint. Also in this case we have that active
stress and active strain do not produce the same stress on
a simple shear.  The non-vanishing components of the
stresses along the simple shear~\eqref{eq:shear} are showed in
Fig.~\ref{fig:stressnostra}.
We notice that the exponential form of the energy amplifies the
differences between the two activation approaches, as already remarked
in~\cite{Rossi2012}.

\section{Decoupled active strain approach}
\label{sec:decoupled}

Active stress and active strain are by far the two most used methods
in the continuum modeling of activation, at least for biological
tissues; however, other approaches can be found in the literature. In
this brief section we study one of them, which is a sort of active
strain applied only to a part of the
energy~\cite{Hernandez13,Pae15}. We will see that such an approach is
equivalent to an active strain.

Let us consider a passive energy of the form
\[
W_\text{pas}=W(\F)+W_\text{aniso}(I_4)
\]
(for instance, if $W$ is assumed to be isotropic, we get the
fiber-reinforced materials introduced in Sect.~\ref{sec:fr}).
Now apply the Kr\"oner-Lee decomposition only to $W_\text{aniso}$,
so that the energy of the active material is given by
\[
W_\text{decoup}=W(\F)+(\det\F_a)W_\text{aniso}(I_4(\C_e)).
\]
We name \emph{decoupled active strain} such an approach. 
Notice that in the full active strain method one should compute
also $W$ on the elastic part $\F_e=\F\F_a^{-1}$, as in~\eqref{eq:Wstrain}.

Let us assume that $\F_a$ can depend on the
deformation gradient $\F$ only through the invariant $I_4$ and that
$\m$ is an eigenvector of $\F_a$ (for instance, if $a$ is a function
only of $I_4$, the active strain~\eqref{eq:Fa} 
satisfies the two assumptions). Then we claim that 
\begin{center}
\emph{the decoupled active
strain is completely equivalent to an active stress approach.} 
\end{center}

Indeed, we prove that there exists a suitable energy density
$W_\text{act}(\sqrt{I_4})$ such that
\[
W_\text{stress}=W_\text{pas}+W_\text{act}=W_\text{decoup}
\]
on \emph{every} deformation. Hence, the two methods produce the same
stress even in the case of simple shear.

Indeed, the two energies coincide if
\begin{equation}
\label{eq:defWa}
W_\text{act}(\sqrt{I_4})=W_\text{aniso}(I_4(\C_e))-W_\text{aniso}(I_4),
\end{equation}
but in general the quantity $I_4(\C_e)$ depends on the whole $\F$ and not
only on $I_4$. However, in the case when $\m$ is an eigenvector of
$\F_a$, it is easy to verify that 
\[
I_4(\C_e)=\frac{I_4}{I_4(\C_a)},
\]
where $\C_a=\F_a^\T\F_a$.
Moreover, we assumed that $\F_a$ is a function only of
$I_4$, hence~\eqref{eq:defWa} is a good definition for
$W_\text{act}(\sqrt{I_4})$. Then, active stress and decoupled active strain give
the same stress on every deformation.

\section{Conclusions}

The present paper shows that the two main approaches to 
activation in Continuum Mechanics, namely active strain and active
stress, give different results on a simple shear deformation even if
they exploit the same passive energy and coincide in the active case on uniaxial deformations along the anisotropy direction. 

We have assumed that the passive material is transversely
isotropic and incompressible. Following the most
widespread constitutive prescriptions in the literature, we have constitutively prescribed either the active strain
tensor $\F_a$ or the active stress tensor $\P_\text{act}$.
In the first case, we have assumed that the active strain is isochoric. We have found a difference
between the two activation models also in the case of a compressible active
strain, namely when $\det\F_a\neq1$, even if the results are not
reported here.

In Sect.~\ref{sec:stressfromstrain} we have considered a hyperelastic
material which is transversely isotropic and incompressible, with a
strain energy density of the form $W(I_1,I_2,I_4, I_5)$. Given an active
strain model with a constant incompressible activation, the active
stress approach has been set up to show the same behaviour on
uniaxial deformations. We have then tested the response of the active
material on a shear deformation. The two activation approaches
coincide if and only if the very restrictive condition~\eqref{eq:Wlx}
holds; moreover, we have showed that the typical energy densities used in
nonlinear elasticity, such as the fiber-reinforced Mooney-Rivlin
energy, do not satisfy the condition. 

A quantitative comparison of the two activation approaches on the
simple shear has been carried out in Sect.~\ref{sec:Fainco} and
Sect.~\ref{sec:energyebi}. In the former we have considered a
fiber-reinforced Mooney-Rivlin material with an isochoric active strain,
while the latter dealt with an energy which is typically used for
the skeletal muscle tissue and where the active stress on the uniaxial
deformation comes from experimental data. Here an active
strain which depends on the deformation had to be taken into account.
In all the cases, it is found that the two activation models do not
coincide on a simple shear deformation.

In Sect.~\ref{sec:decoupled} we have discussed a slightly different
approach to active strain, sometimes used in the biomechanical
literature related to muscles, which we have named decoupled active strain.
It turns out that it is completely equivalent to the active stress, at
least if the anisotropic part of the energy depends only on $I_4$.

Our results may be useful in developing new models of anisotropic
active materials: indeed, from Figs.~\ref{fig:WincoFaincocomp},
\ref{fig:WincoFaincodifferenze},
\ref{fig:stressnostra}, it is clear that experimental data on the
stress-stretch response on uniaxial deformations are not enough to
characterize the behavior of the active material. In order to construct a
more realistic model, reliable on other classes of deformation, it
is necessary to perform further experiments, for example on simple
shears. 
Notice that there are a few
experimental works considering deformation modes other than the uniaxial traction but, as far as we
know, they study only the passive case (see for instance~\cite{Morrow2010}). 

 \section*{Acknowledgement}
The authors thank the anonymous reviewers for their comments and suggestions.

  This work has been partially supported by National Group of
  Mathematical Physics (GNFM-INdAM).

\end{document}